\documentclass[apj,numberedappendix]{emulateapj}
\usepackage{epsfig}
\usepackage{amsmath}
\usepackage{enumerate}
\usepackage{natbib}
\usepackage{hyperref}
\usepackage{ulem}

\usepackage{color}

\def\sT{\sigma_{\rm T}}
\def\beq{\begin{equation}}
\def\eeq{\end{equation}}
\def\Ep{E_{\rm pk}}

\def\Sect{Section}
\def\Sects{Sections}
\def\Eq{Equation}
\def\Eqs{Equations}
\def\Fig{Figure}
\def\texp{t_{\rm exp}}

\def\ep{\epsilon}

\def\vbw{v_{\rm bw}}
\def\Gbw{\Gamma}
\def\Tp{T_{p}}
\def\Tb{T_{\rm GRB}}

\def\tobs{t_{\rm obs}}
\def\Rload{R_{\rm load}}

\def\EIC{E_{\rm IC}}
\def\Lej{L_{\rm ej}}
\def\Lb{L_{\rm GRB}}
\def\Eb{E_{\rm GRB}}
\def\ge{\gamma_e}

\def\xiacc{\xi_{\rm acc}}
\def\Zacc{Z_{\rm acc}}
\def\Grel{\Gamma_{\rm rel}}
\def\Gsh{\Gamma_{\rm FS}}
\def\gpre{\gamma}
\def\bbw{\beta_{\rm bw}}
\def\bpre{\beta}
\def\Gej{\Gamma_{\rm ej}}
 
\def\RFS{R_{\rm FS}}

\def\ginj{\gamma_{\rm inj}}

\def\Et{E_t}
\def\M{{\cal M}}
\def\Ediss{E_{\rm diss}}
\def\Eflash{E_{\rm flash}}

\def\x{\varpi}
\def\xsc{\varpi_{\rm sc}}
\def\tlab{t_{\rm lab}}
\def\xabs{\varpi_{\rm abs}}

\def\eff{\varepsilon_{\rm rad}}
\def\tFS{t_{\rm FS}}
\def\xFS{\varpi_{\rm FS}}

\def\EKN{E_{\rm KN}}

\def\tc{t_{\rm IC}}
\def\ep{\epsilon_{\rm pk}}
\def\Rsc{R_{\rm IC}}
\def\thsc{\theta_{\rm IC}}
\def\thesc{\theta_{\rm esc}}
\def\Lp{L_{\epsilon}^{\rm pk}}
\def\epsun{\epsilon_0}
\def\UT{U_{\rm T}}
\def\fT{f_{\rm T}}
\def\go{\gamma_{\rm opt}}
\def\fsyn{f_{\rm syn}}
\def\Lo{L_{\rm opt}}
\def\gth{\gamma_{\rm th}}
\def\Ldiss{L_{\rm diss}}
\def\Uup{U_{\rm up}}

\def\Lpm{L_\pm}

\def\Lflash{L_{\rm flash}}
\def\gh{\gamma_{\rm heat}}

\def\taugg{\tau_{\gamma\gamma}}
\def\sigmagg{\sigma_{\gamma\gamma}}

\def\epss{\epsilon_t}
\def\epscm{\epsilon_{\rm cm}}
\def\epstrh{\epsilon_{\rm thr}}
\def\epsp{\epsilon_{\rm pk}}
\def\sigmat{\sigma_{\rm T}}

\def\epssc{\epsilon_{\rm sc}}
\def\Isc{I_{\rm sc}}

\def\me{m_{\rm e}}

\def\Zpm{Z_{\pm}}
\def\Zacc{Z_{\rm acc}}
\def\xiacc{\xi_{\rm acc}}

\def\musc{\mu_{\rm sc}}

\def\epsB{\varepsilon_{\rm B}}
\def\Grel{\Gamma_{\rm rel}}

\def\epse{\varepsilon_{\rm e}}

\newbox\grsign \setbox\grsign=\hbox{$>$} \newdimen\grdimen \grdimen=\ht\grsign
\newbox\simlessbox \newbox\simgreatbox \newbox\simpropbox
\setbox\simgreatbox=\hbox{\raise.5ex\hbox{$>$}\llap
     {\lower.5ex\hbox{$\sim$}}}\ht1=\grdimen\dp1=0pt
\setbox\simlessbox=\hbox{\raise.5ex\hbox{$<$}\llap
     {\lower.5ex\hbox{$\sim$}}}\ht2=\grdimen\dp2=0pt
\setbox\simpropbox=\hbox{\raise.5ex\hbox{$\propto$}\llap
     {\lower.5ex\hbox{$\sim$}}}\ht2=\grdimen\dp2=0pt
\def\simgt{\mathrel{\copy\simgreatbox}}
\def\simlt{\mathrel{\copy\simlessbox}}

\shorttitle{GeV emission in gamma-ray bursts}
\shortauthors{A. M. Beloborodov, R. Hasco\"et, I. Vurm}

\begin{document}

\title{On the origin of GeV emission in gamma-ray bursts}

\author{Andrei M. Beloborodov\altaffilmark{1}, Romain Hasco\"et\altaffilmark{1} and Indrek Vurm\altaffilmark{1,2}}
\affil{$^1$Physics Department and Columbia Astrophysics Laboratory, Columbia University, 538 West 120th Street, New York, NY 10027, USA; amb@phys.columbia.edu \\
$^2$Tartu Observatory, T\~{o}ravere 61602, Tartumaa, Estonia\\
}

\label{firstpage}
\begin{abstract}
The most common progenitors of gamma-ray bursts (GRBs) are massive 
stars with strong stellar winds. We show that the GRB blast wave in the wind
should emit a bright GeV flash. 
It is produced by inverse Compton 
cooling of the thermal plasma behind the forward shock wave.
The main part of the flash is shaped by
scattering of the prompt MeV radiation (emitted at smaller radii) which streams 
through the external blast wave. 
The inverse-Compton flash is bright due to the huge $e^\pm$ enrichment 
of the external medium. 
At late times, the blast wave switches to normal synchrotron-self-Compton cooling.
The mechanism is demonstrated by a detailed transfer simulation.
The observed prompt MeV radiation is taken as an input of the simulation; 
we use GRB~080916C as an example.
The result reproduces the GeV flash observed by the {\it Fermi} telescope.
It explains the delayed onset, the steep rise, the peak flux, the time of the peak,
the long smooth decline, and the spectral slope of GeV emission.
The wind density required to reproduce all these features is typical of 
Wolf-Rayet stars. Our simulation predicts strong TeV emission 1~min after the 
burst trigger; then a cutoff in the observed high-energy spectrum is expected 
from absorption by extragalactic background light.
In addition, a bright optical counterpart of the GeV flash is predicted for plausible values of the 
magnetic field; such a double (optical+GeV) flash has been observed in GRB~130427A.
\end{abstract}


%############################################################

\section{Introduction}

The luminosities of gamma-ray bursts (GRBs) peak in the soft gamma-ray
band around 1~MeV (e.g. \citealt{goldstein_2012}). Observations by the Large Area Telescope
(LAT) onboard the {\it Fermi} satellite \citep{atwood_2009} show that some GRBs
also give rise to a longer GeV flash, with a distinct light curve \citep{lat_2013}.
The energy emitted in the GeV band is smaller than
that of the  main (``prompt'') MeV radiation,
typically by a factor $\sim 10$. Nevertheless, as we argue in this paper,
it can play a key role for
understanding the nature of GRB explosions and their progenitors.

The GeV flash can shed light on the explosion picture only if 
its radiative mechanism is identified with some confidence. Ideally, one would hope for
a model that reproduces the observed light curve and spectrum 
from a first-principle calculation.
In search of such a model, one can consider various possibilities such as
synchrotron emission from the blast wave \citep{zou_2009, kumar_2009, ghisellini_2010},
hadronic processes (e.g. \citealt{asano_2009,razzaque_2010}), or inverse 
Compton emission from internal shocks (e.g. \citealt{bosnjak_2009,toma_2011}).
None of the proposed models, however, predict the observed light curve,
and most models invoke extreme parameters (low external density 
and magnetic fields, or a huge explosion energy).
The synchrotron mechanism of GeV emission is problematic as
it requires extreme particle acceleration; 
even under most favorable conditions it cannot explain the observed 
spectrum which extends to 100~GeV (e.g. 
\citealt{piran_2010, sironi_2013, wang_2013}).

The radiative process capable of producing the observed flash is inverse 
Compton (IC) scattering; the seed photons for IC scattering can be provided 
by the prompt GRB or its afterglow radiation.
In particular, \citet{beloborodov_2005} suggested that GRBs should be accompanied 
by GeV flashes due to 
IC scattering of the prompt MeV radiation streaming through the 
external blast wave. 
Observations by {\it Fermi} LAT provide support to this picture: 

(1) In practically all GRBs detected by {\it Fermi} LAT (except a few cases with 
poor photon statistics) the peak of the GeV flash
overlaps with the prompt MeV radiation \citep{lat_2013}.
The overlap implies that the GeV source experiences Compton cooling
by the prompt MeV radiation (keV radiation in the rest frame of the source).

(2) The GeV flash has a distinct light curve, different from the prompt
MeV burst. It quickly rises and then
shows a long monotonic decay, which lasts significantly longer
than the prompt MeV emission. This is expected if the GeV flash
is produced by the external blast wave.
The blast wave has a larger radius and moves with a smaller Lorentz factor
compared with the source of the prompt burst, and hence its emission can be
spread over longer observational times.

(3) The onset of GeV emission is slightly delayed with respect to the
beginning of the prompt MeV burst.
The arrival time of photons emitted by the blast wave at radius $R$
is roughly given by
\beq
\label{eq:tobs}
    \tobs\sim(1+z)\left(\frac{R}{\vbw}-\frac{R}{c}\right)
            \approx (1+z)\,\frac{R}{2\Gbw^2c}.
\eeq
Here $\tobs$ is measured by the clock of a distant observer
since the first light signal from the beginning of the explosion, 
$\Gbw=(1-\vbw^2/c^2)^{-1/2}\gg 1$ is the Lorentz factor of the blast
wave, and $z$ is the cosmological redshift.
The delay in the onset of GeV emission is expected if the blast-wave
luminosity is suppressed at small radii. It equals the time it takes the explosion
to reach the radius where it becomes a bright GeV source,
which is typically a few seconds.\footnote{The prompt burst 
is emitted at a much smaller radius $R_{\rm MeV}$, with a Lorentz 
factor $\Gej\simgt\Gbw$. Therefore, its delay
$\sim (1+z)(R_{\rm MeV}/2\Gej^2c)$ is much smaller.}

However, any model associating the GeV flash with the 
external blast wave faces the following puzzle.
Many observed GeV flashes reach the peak and start to decay at time $\Tp$
much shorter than the duration of the prompt MeV burst, $\Tb$.
For example, GRB~080916C has $\Tp\sim 0.1\Tb$  \citep{abdo_2009}.
Why would the peak of blast-wave radiation
be much shorter than the prompt burst itself?
Consider the standard model where $\Tb$ corresponds to the duration of the
ultra-relativistic ejecta that emits the prompt burst.
Then $c\Tb(1+z)^{-1}$ is a measure of the ejecta thickness.
The ejecta energy is transferred to the blast wave through the
reverse shock, which may be relativistic and can cross the ejecta as quickly 
as $T_{\rm cross}\sim \Tb$
(in observer time). The ejecta cannot transfer its energy at $\tobs\ll\Tb$,
as this would require a superluminal motion of the reverse shock,
and hence the self-similar deceleration of the blast wave
should not begin until $\tobs\sim \Tb$. Then the GeV flash is not
expected to decay until $\tobs\sim \Tb$ (e.g. \citealt{gao_2009}; 
\citealt{he_2011, maxham_2011}).
The problem becomes even more severe in 
explosion models with a non-relativistic reverse shock; then the 
deceleration/decay stage is not expected until $\tobs\gg\Tb$.

This puzzle is resolved by the fact that {\it the blast 
wave propagates in a medium with a quickly changing composition.}
As discussed in detail below, at radii $R\simlt 10^{16}$~cm the medium 
is extremely rich in $e^\pm$ pairs, with $Z_\pm\simgt 10^4$ pairs per ion. 
Pairs are inevitably produced by the prompt MeV radiation propagating
ahead of the blast wave
(\citealt{thompson_2000}; \citealt{meszaros_2001}; \citealt{beloborodov_2002}, 
hereafter B02; \citealt{kumar_2004}).
The huge number of the prompt MeV photons ($N_{\rm MeV}\sim 10^{60}$
in isotropic equivalent for the brightest GRBs) implies exponential pair 
creation in a static optically thin medium. In addition, radiation
exerts a strong force and significantly accelerates the external medium, 
which affects the strength of the forward shock
and the evolution of its temperature.

We show in this paper that the forward shock propagating in the 
pre-accelerated pair-enriched medium 
is an extremely efficient producer of GeV emission, regardless of the
details of the shock microphysics and its efficiency in nonthermal
particle acceleration. This provides a robust mechanism for a GeV flash.
As the blast wave expands to larger radii where $Z_\pm$ is reduced, its
GeV luminosity decreases.

A possible role of $e^\pm$ loading for GeV emission
was previously conjectured by \citet{ghisellini_2010}, although their
scenario differs from the model presented here.
\citet{ghisellini_2010} assumed that the blast wave enters the self-similar 
deceleration stage in the pair-dominated zone and continues to radiate
with the pair-assisted efficiency close to 100\%.
They explained the observed decline of GeV emission
by the steep reduction of the dissipation power in the decelerating blast wave.
As discussed above, the problem of this picture is
that the self-similar deceleration should not begin until $\tobs\sim\Tb$
while the observed decline in many LAT bursts starts at $\Tp\ll\Tb$.\footnote{A
related technical remark: \citet{ghisellini_2010} used energy
of the entire burst in the estimate of the pair-loading effect. 
In fact, when the GeV flash peaks, only a fraction $\sim\Tp/\Tb$
of the prompt burst is ahead of the blast wave,
and $\Eb$ contributing to its pair loading 
is reduced by a factor of $\sim \Tp/\Tb$.}
Another difference concerns the emission mechanism: \citet{ghisellini_2010} 
associated GeV photons with synchrotron emission from nonthermal particles.
We find that the GeV flash is produced by inverse Compton scattering 
of the prompt radiation by the thermal plasma behind the forward shock.

In this paper, we study explosions in the wind medium expected from a 
massive progenitor (e.g. \citealt{chevalier_1999}).   
We consider a Wolf-Rayet star with a typical mass-loss 
rate $\dot{M}\approx 10^{-5}M_\odot$~yr$^{-1}$, which produces a wind with 
density profile $\rho\propto R^{-2}$.
We calculate the dynamics, $e^\pm$ density, and temperature of the blast
wave and show that it must generate an inverse-Compton pair-dominated 
flash in the GeV band. Its light curve and spectrum can be calculated 
from first principles, using a direct simulation of radiative transfer.
  
Preliminary estimates explaining the proposed mechanism are presented in 
\Sect~2. Then in \Sects~3 and 4 we describe the setup of our detailed 
calculations. The results are described in \Sect~5 using
GRB~080916C as an example. In \Sect~6, we present analytical estimates for 
photon-photon ($\gamma$-$\gamma$) opacity. Then, in \Sect~7, we 
discuss the expected synchrotron emission from the pair-loaded blast wave, 
and find that, in a broad range of the magnetization parameter
$\epsB$, the GeV flash is accompanied by a bright and brief 
optical flash. In \Sect~8 we estimate the effect of the GeV flash on the 
external medium. Our results are summarized and discussed in \Sect~9.

%############################################################

\section{Preliminary estimates}

\subsection{Number of GeV photons in the flash}

Consider a blast wave that sweeps up the external medium.
Let $\ginj$ be the mean (thermal) Lorentz factor of hot electrons immediately
behind the forward shock, and $\Gamma$ be the bulk Lorentz factor of the shocked fluid. 
Subscript ``inj'' in $\ginj$ stands for ``injection'' --- the hot plasma 
is injected at the shock front and cools down behind it.

The plasma is Compton cooled by the
prompt GRB radiation that gradually leaks out of the explosion ejecta and
streams radially through the external blast
wave.\footnote{
The prompt photons are assumed to be emitted at a small radius 
$R_{\rm MeV}\ll R$, and their angles with respect to the radial 
direction are $\theta\sim(R_{\rm MeV}/R)\,\Gej^{-1}\ll \Gbw^{-1}$.
}
Let $\Et\sim 1$~MeV be the characteristic energy of the prompt photons in 
the lab frame; they serve as targets for inverse Compton (IC) scattering.
Their energies in the fluid frame are
\beq
  \Et^\prime=\frac{\Et}{2\Gamma}.
\eeq
The hot electrons injected at the shock front lose
energy by upscattering the target photons.
The typical energy of upscattered photons in the fluid frame is
$\EIC^\prime\sim\ge^2\Et^\prime$ (assuming Thomson scattering).
The corresponding energy of IC photons in the lab frame is
$\EIC\approx (2/3)\Gamma\EIC^\prime$, 
\beq
\label{eq:EIC}
   \EIC\sim \frac{1}{3}\ge^2\Et.
\eeq
One can see that GeV photons are generated when 
\begin{equation}
\label{eq:ge}
  \ge\sim 50\,\left(\frac{\EIC}{1{\rm ~GeV}}\right)^{1/2}
                    \left(\frac{\Et}{{\rm 1~MeV}}\right)^{-1/2}.
\end{equation}
Then one can verify that the scattering is in the Thomson regime, 
$\ge\Et^\prime/m_ec^2<1$, although moderate Klein-Nishina corrections 
are beginning to appear at these energies.

As the electron injected with $\ge=\ginj$ cools down, it produces IC photons with
decreasing  $\EIC\propto \ge^2$.
Their number near a given energy $\EIC$ may be estimated as
\beq
\label{eq:M}
   \M\sim \frac{\ge m_ec^2}{\EIC^\prime}
       \sim \frac{m_ec^2}{(\Et^\prime\EIC^\prime)^{1/2}}
       \sim \frac{\Gamma m_ec^2}{(\Et \EIC)^{1/2}}.
\eeq
The multiplicity of photons with $\EIC\sim 1$~GeV produced by an electron with 
$\ginj\gg 50$ is $\M\sim \Gamma/60$.

The number of GeV photons produced by the shock wave is
\beq
\label{eq:NGeV}
    N_{\rm GeV}\sim \M N_{\pm},
\eeq
where $N_{\pm}$ is the number of electrons/positrons swept-up by the shock, 
proportional to the total swept-up mass $m$,
\beq
\label{eq:Ne}
   N_{\pm}=Z_\pm\,N_p, \qquad N_p=\frac{m}{\mu_e m_p}.
\eeq
Here $Z_\pm$ is the pair loading factor of the external medium,
$N_p$ is the number of swept-up protons, and
$\mu_e$ depends on the chemical composition of the medium;
$\mu_e=1$ for hydrogen and $\mu_e=2$ for heavier elements.

The medium is expected to be a wind from a massive progenitor,
which is losing mass before the explosion with a rate
$\dot{M}$. The mass of the wind medium contained in a sphere of radius $R$ is
given by
\beq
    m(R)=\frac{\dot{M} R}{w},
\eeq
where $w$ is the wind velocity. The likely GRB progenitors are Wolf-Rayet
stars, whose observed winds have typical 
$\dot{M}\sim 10^{-5}M_\odot$~yr$^{-1}$,
$w\sim 2\times 10^8$~cm~s$^{-1}$, and $\mu_e\approx 2$ 
(e.g. \citealt{hamann_1995, lamers_1999, crowther_2007}).
This gives
\beq
    N_p\sim 10^{52} R_{16} \dot{M}_{-5},
\eeq
where $\dot{M}_{-5}=\dot{M}/10^{-5} M_\odot{\rm ~yr}^{-1}$
and $R_{16}=R/10^{16}$cm.
  
The value of $Z_\pm$ can be exactly calculated using the observed 
luminosity and spectrum of the prompt GRB (\Sect~3.1); it has enormous 
values $Z_\pm \sim 10^{5}$ at the early stages of blast-wave expansion
and then steeply decreases with radius.
In particular, for GRB~080916C we
will show below that the GeV 
flash peaks at a well-defined radius $R_p\approx 10^{16}$~cm where 
$Z_\pm\sim 10^4$.
 \Eqs~(\ref{eq:NGeV}) and (\ref{eq:Ne}) with $Z_\pm\sim 10^4$ give
a rough estimate for the number of GeV photons,
\beq
   N_{\rm GeV}\sim 10^{57},
\eeq
which is close to the isotropic equivalent of the bright GeV flashes
observed by LAT. The high density of the progenitor wind and the huge 
pair enrichment is what makes the inverse Compton mechanism capable 
of emitting a bright 
flash; models neglecting 
pair creation would fall far short in $N_{\rm GeV}$.

Note that the prompt GRB radiation plays a key role for the 
GeV flash in two ways: (1) it provides
target photons for IC scattering and 
(2) its interaction with the external medium ahead of the shock ensures
the $e^\pm$ enrichment of the medium.
The $e^\pm$ pairs radiating GeV photons behind the shock are
created by the prompt MeV photons propagating ahead of the shock.
The total number of the prompt photons in a burst like GRB~080916C is huge,
$N_{\rm MeV}\sim 10^{60}$ (isotropic equivalent). Almost all these photons pass
through the external medium unaffected, as the medium is optically thin.
A small fraction of photons get scattered and converted to $e^\pm$ pairs,  
so the number of created pairs $N_\pm\ll N_{\rm MeV}$. However, $N_\pm$ 
greatly exceeds $N_p$, by the factor $Z_\pm\gg 1$.

\subsection{Radiative efficiency in the GeV band}

As will be demonstrated with detailed calculations below, the external blast wave
inevitably passes through a stage with the pair-loading factor $Z_\pm\sim 10^4$
and pre-acceleration Lorentz factor $\gpre\sim 10$. It is an extremely efficient 
producer of GeV emission at this stage.
Three factors contribute to the high efficiency:
 
(1) The high pair-loading factor $Z_\pm\sim 10^4>m_p/m_e$ guarantees 
that most of the shock-dissipated energy is given to leptons.

(2) At this stage, the shock-heated pairs have the thermal Lorentz factor 
$\ginj\sim\Gbw/\gpre\sim 50$, so their IC cooling produces emission in the GeV 
band according to \Eq~(\ref{eq:EIC}). The
relatively low value of $\ginj$ is a result of pair loading and pre-acceleration of 
the external medium. 
Note that pre-acceleration reduces the strength of the forward shock:
the fluid Lorentz factor jumps at
the shock front from $\gpre\sim 10$ to $\Gbw$, which corresponds to 
electron heating to $\ginj\sim\Gbw/\gpre$.\footnote{
    Energy transfer from the shocked ions to electrons is unable to 
    significantly increase $\ginj$ in the medium with $Z_\pm\sim 10^4$,
    since the ion abundance is smaller than $m_e/m_p$.
    This effect can, however, become significant soon after the peak of the flash, 
    as $Z_\pm$ decreases.
    }

(3) Inverse Compton cooling of the shocked pairs is fast, so they efficiently 
radiate their energy. The cooling timescale of isotropic electrons with Lorentz factor 
$\ge$ in the fluid frame is given by
\beq
\label{eq:tc}
   \tc^\prime=\frac{3m_ec}{4\,\sT U^\prime \ge},
\eeq
where $U^\prime=(2\Gbw)^{-2}U$ is the energy density of the collimated 
prompt radiation in the fluid frame, and $U=\Lb/4\pi R^2 c$. The cooling 
timescale should be compared with the expansion timescale of the blast
wave, $\texp^\prime=R/c\Gbw$,
\beq
   \frac{\tc^\prime}{\texp^\prime} = 
       \frac{12\pi\,m_ec^3 R\,\Gbw^3}{\sT \Lb\ge} 
     \approx 
       \frac{2}{\ge}\, 
   R_{16} L_{54}^{-1}\left(\frac{\Gbw}{500}\right)^{3},
\eeq
which gives $\tc^\prime<\texp^\prime$ 
for $\ge\gg 1$. Compton cooling is fast for electrons 
emitting in the GeV band, $\ge\simgt 50$. Electrons with $\ge\gg 10^2$ scatter
photons with a smaller rate due to the Klein-Nishina correction, but their cooling is still fast.

\subsection{Lorentz factor of the blast wave and arrival time of GeV photons}
\label{subsect_gbw}

The arrival time of IC photons emitted at radius $R_p\sim 10^{16}$~cm
(peak of the GeV flash) depends on the Lorentz factor of the blast wave,
$\Gbw$, according to \Eq~(\ref{eq:tobs}). Note that $R_p$ 
can be significantly smaller than the radius where the blast wave enters
the self-similar deceleration. At this early stage, the blast-wave material is
sandwiched between the forward and reverse shocks, and its Lorentz
factor $\Gbw$ is regulated by the ram pressures in the two shocks, $P_f$
and $P_r$.

An estimate for $\Gbw$ may be obtained assuming pressure balance 
$P_f\sim P_r$. A convenient approximation for the shock pressure is given 
by \citep{beloborodov_2006},
\beq
\label{eq:P}
  P=\frac{4}{3}\left(\Grel^2-1\right)\Uup,
\eeq
where $\Grel$ is the relative Lorentz factor of the upstream and downstream,
and $\Uup= \gamma(1+\beta) \gh\rho c^2$ is the proper energy density of 
the upstream fluid; $\gh-1$ is a measure of upstream heat relative to the rest 
mass, and we took into account that the pre-accelerated external medium is 
compressed by the factor of $\gamma(1+\beta)$
as required by the continuity equation (B02).
This gives,
\beq
\label{eq:Pf}
   P_f\approx  \frac{4}{3}
  \frac{\Gbw^2 \gh\,\rho c^2}{\gamma(1+\beta)} 
     \left(1+\frac{Z_\pm m_e}{\mu_em_p}\right),
\eeq
where we used $\Grel\approx \Gbw/\gpre(1+\bpre)\gg 1$. 
In the absence of pre-heating and pre-acceleration ($\gh=\gpre=1$) and 
moderate pair loading ($Z_\pm\ll m_p/m_e$), \Eq~(\ref{eq:Pf}) reduces to the 
standard relation $P_f=(4/3)\Gbw^2\rho c^2$.

For the reverse shock one can use \Eq~(\ref{eq:P}) with
$\Grel\approx (1/2)(\Gej/\Gbw+\Gbw/\Gej)$ and $\Uup=\rho_{\rm ej}c^2$,
\beq
\label{eq:Pr}
  P_r\approx \frac{1}{3}\left(\frac{\Gej}{\Gbw}-\frac{\Gbw}{\Gej}\right)^2
                                  \rho_{\rm ej} c^2,                                 
\eeq
where $\rho_{\rm ej}$ and $\Gej$ are the fluid mass density and Lorentz factor 
of the ejecta. Then the pressure balance $P_f\sim P_r$ gives
\beq
\label{eq:Gbw_gen}
   \Gbw\approx \Gej\left\{1+2\Gej^2\left[\frac{4\pi A c^3
     \gh
    (1+Z_\pm m_e/\mu_e m_p)}{\Lej \gpre(1+\bpre)}\right]^{1/2}\right\}^{-1/2}.
\eeq
Here $\Lej=4\pi R^2 \Gej^2\rho_{\rm ej} c^3$ is the kinetic power of the ejecta
(isotropic equivalent) and we used the external density profile $\rho=A R^{-2}$ 
where $A\equiv \dot{M}/4\pi w\sim 10^{11}-10^{12}$~g~cm$^{-1}$.
In the case of a relativistic reverse shock, $\Gej^2\gg\Gbw^2$, the expression
for $\Gbw$ simplifies and becomes independent of $\Gej$,
\beq
\label{eq:Gbw}
    \Gbw\approx\left[\frac{\Lej\, \gpre(1+\bpre)}{16\pi A c^3
      \gh
                  (1+Z_\pm m_e/\mu_e m_p)}\right]^{1/4}.
\eeq
This equation
gives $\Gbw\sim 500$ for the parameters of GRB~080916C
discussed in this paper.
Note that $\gpre$ is determined by the force exerted by the prompt radiation 
front ahead of the blast wave. Our numerical calculations give $\gpre\sim 10$, 
$Z_\pm\sim 10^4$, and $\gh\approx 1$
at the peak radius of the GeV flash, $R_p\approx 10^{16}$~cm (see \Sect~5).

Using \Eqs~(\ref{eq:tobs}) and (\ref{eq:Gbw}), one finds the arrival time
of the peak of the flash $\tobs\sim 1-10$~s, which is consistent with
observations.
The detailed calculations presented below will give a more accurate
estimate for the arrival time of the peak. We will also calculate the light 
curve of the GeV flash and show that its decay
after the peak extends over much longer times.

\subsection{Energy dissipated in the forward shock}

As a final check, let us estimate the energy dissipated in the forward
shock near the radius $R_p\sim 10^{16}$~cm. Since
most of the dissipated energy $\Ediss$ is radiated in GeV photons,
one expects a GeV flash of energy $\Eflash\sim\Ediss$.

The dissipation rate in the forward shock is approximately given by,
\beq
   \Ldiss\approx 4\pi R^2 (3P_f) \Gbw^2 c\sim 4\pi R^2 (3P_r) \Gbw^2 c \sim \Lej,
\eeq
where we used \Eq~(\ref{eq:Pr}) and assumed $\Gej \gg \Gbw$. The ejecta
power $\Lej$ is comparable to or larger than the
observed luminosity of the prompt GRB, $\Lb$, depending on the
prompt emission efficiency  $\eff$,
\beq
  \frac{\Lej}{\Lb}=\frac{1-\eff}{\eff}.
\eeq
The peak luminosity of the flash $L_{\rm flash}\simlt \Lej$ is  
comparable to $\Lb$ that is observed {\it before} the peak of the flash.
We will confirm this result with more detailed calculations below.

\subsection{Summary}

As the blast wave passes through the radius $R_p\sim 10^{16}$~cm
where $\gpre\sim 10$, the shock wave radiates most 
of the dissipated energy in the GeV band, and the emitted radiation arrives at 
$\tobs\sim 1-10$~s. This defines the peak of the GeV flash. Below we present 
detailed calculations that will give the light curve and spectrum of the flash,
before and after the peak.

%############################################################

\section{Shock wave in pair-loaded medium}

\subsection{Pair loading}

The prompt MeV radiation is nearly perfectly beamed in the radial direction
in the blast-wave region, as it is emitted at much smaller radii.
Those prompt photons that have already
overtaken the forward shock propagate
in the external medium, which has not yet learned about the explosion.
Some of these photons scatter off the ambient medium.
Only a small fraction of photons are scattered (the medium is
optically thin), however this fraction translates into a huge number of scattered
photons {\it per ambient electron}.
Many of these photons quickly convert to $e^\pm$ pairs.
The conversion occurs because the scattered photons
have large angles with respect to the primary (collimated) GRB radiation,
and the large angle lowers the energy threshold for the $\gamma$-$\gamma$
reaction with the beam, $\gamma+\gamma\rightarrow e^++e^-$.

The created pairs also scatter the prompt photons, which leads to
exponential $e^\pm$ creation and a huge enhancement of the
electron density ahead of the forward shock, by a factor $Z_\pm$
exceeding $10^4$ (B02).
The $e^\pm$ loading factor $Z_\pm=n_\pm/n_0\gg 1$ at radii $R<\Rload$,
where
\beq
\label{eq:Rload}
   \Rload\approx 10^{17} E_{\rm GRB,54}^{1/2} {\rm ~cm},
\eeq
and $\Eb$ is the isotropic equivalent of the prompt GRB energy
{\it ahead of the forward shock.}

The main dimensionless parameter that controls $Z_\pm$ at the forward shock 
is proportional to the column density of the GRB radiation ahead of the shock,
\begin{equation}
\label{eq:xi}
  \xi=\frac{\sT}{m_ec^2}\,\frac{\Eb}{4\pi R^2}=650\,E_{\rm GRB,54}R_{16}^{-2}.
\end{equation}
At observer times $\tobs\ll\Tb$, $\Eb$ ahead of the shock is a fraction of the 
total prompt GRB energy (most of which is still behind the shock).
The pair loading factor $Z_\pm(\xi)$ and the pre-acceleration Lorentz factor 
$\gamma(\xi)$ depend only on the prompt radiation field and not on the density 
of the ambient medium (B02). 

We have extended the calculations of B02 in two ways: (1)
B02 assumed a typical prompt GRB spectrum that peaks at $\Ep=m_ec^2$ 
while the bright bursts detected by LAT have higher than average $\Ep$. 
We have extended the model to bursts with high $\Ep\sim 1-10$~MeV.    
(2) B02 used the ``cold approximation'' assuming that the loaded 
$e^\pm$ pairs are quickly cooled to a non-relativistic temperature, so 
that the scattering plasma may be assumed to be cold. This approximation
is accurate only for bursts with $\Ep\ll 1$~MeV. We have relaxed the cold 
approximation and included the thermal motions of pairs in our simulations.

We performed our calculations for 
the prompt radiation with a broken power-law spectrum, whose
    spectral luminosity is given by
\beq
\label{eq:LE}
  L_E=L_E^{\rm pk} \times \left\{\begin{array}{ll}
         (E/\Ep)^{-\alpha_1},  & E<\Ep \\
         (E/\Ep)^{-\alpha_2},  & E>\Ep 
                                       \end{array}
                               \right.
\eeq
As a first test, we ran our code using the cold approximation and found excellent 
agreement with Figures~1-3 in B02. Note that \Eq~(4) in B02 misses 
the factor $d\epsilon/d\epsilon_{\rm sc}$ 
which should have canceled
the factor of $(1+\beta)^{-1}$ in his \Eqs~(42)
and (43). However, the numerical results in B02 are based on the 
correct equations, the missing factor $d\epsilon/d\epsilon_{\rm sc}$ being a 
misprint that propagated to \Eqs~(42) and (43).
   
\begin{figure*}[t]
\begin{tabular}{cc}
\hspace*{0.5cm}
\includegraphics[trim = 0cm 0cm 3cm 0cm, width=0.45\textwidth]{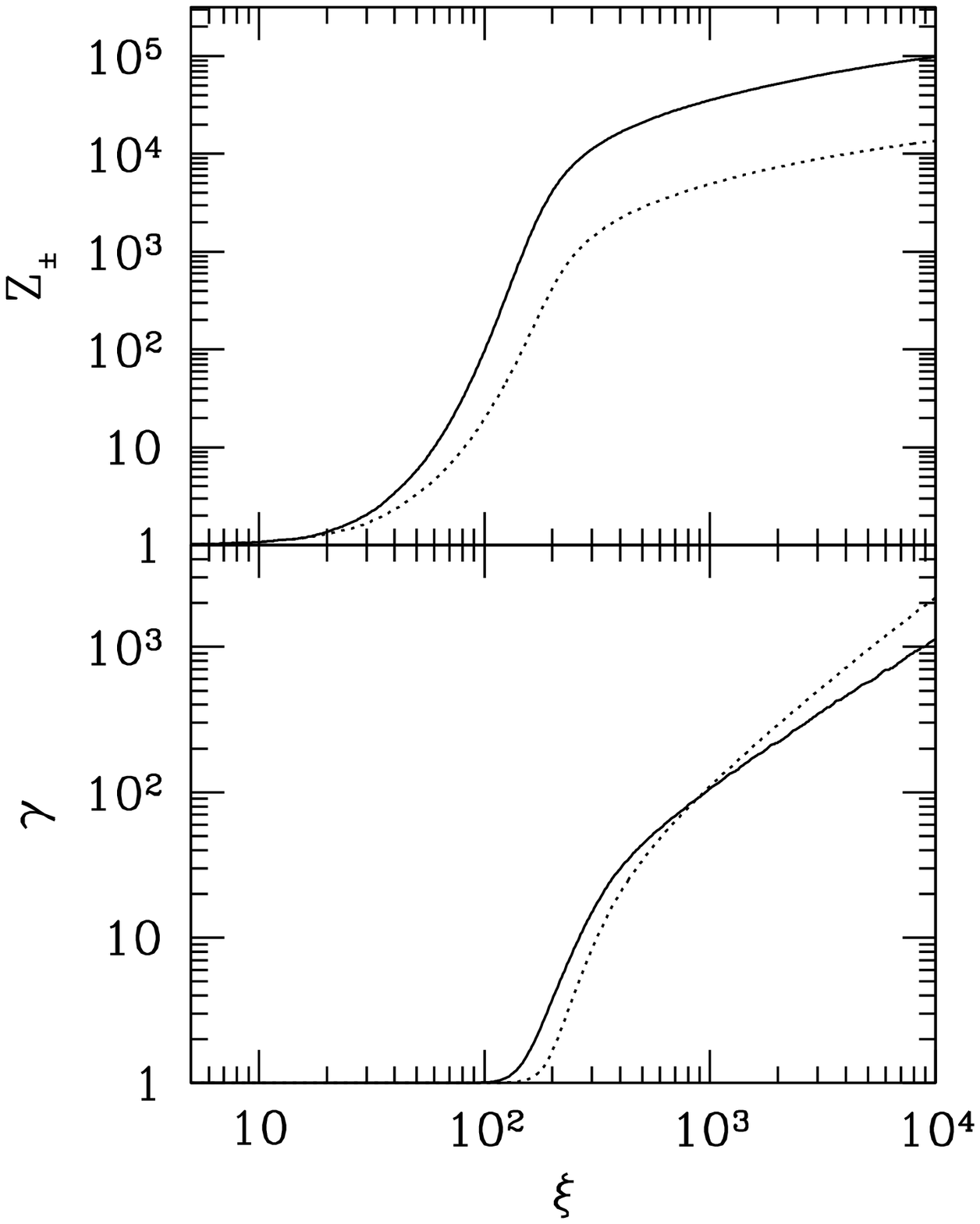} & 
\includegraphics[trim = 2cm 0cm 1cm 0cm, width=0.45\textwidth]{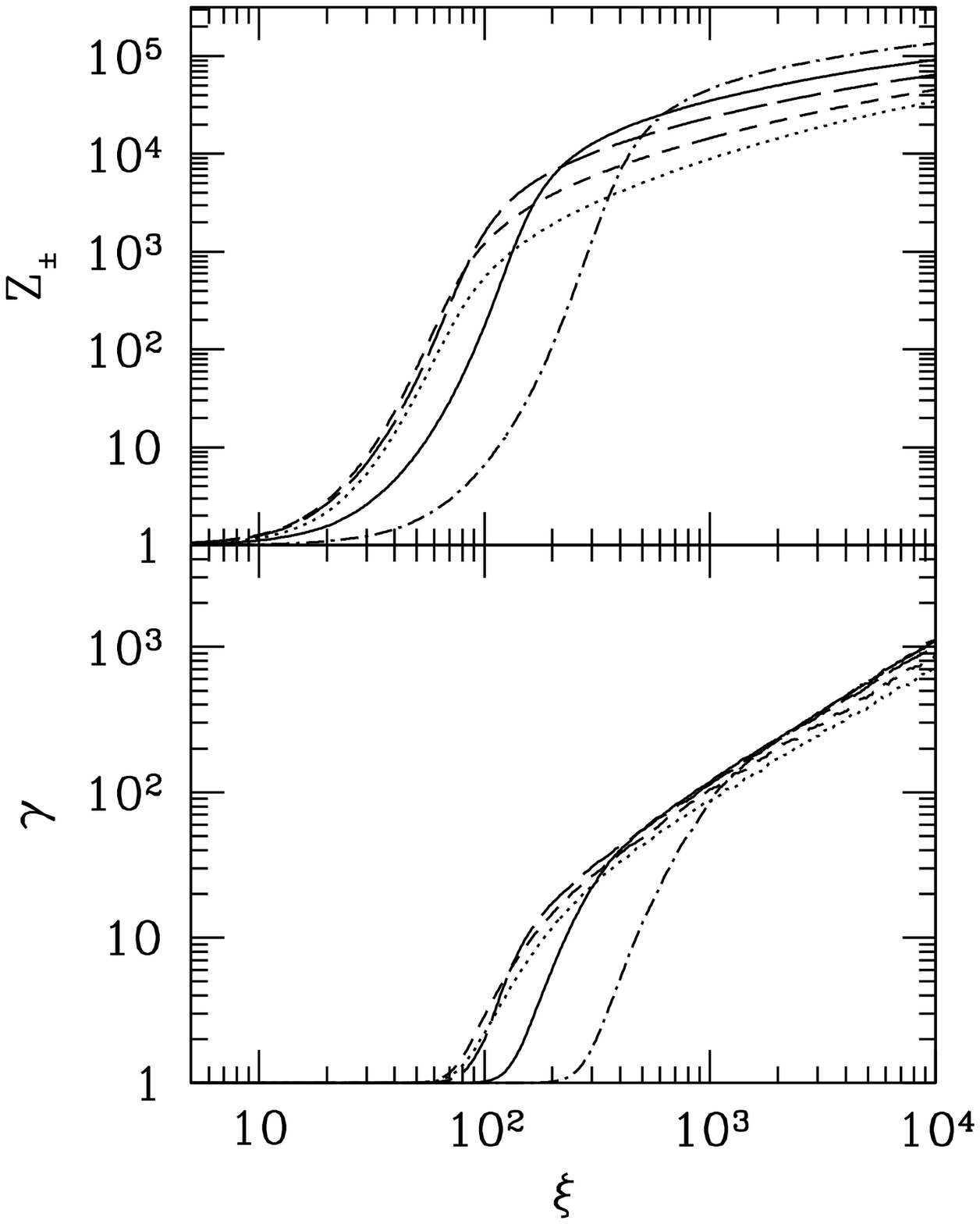}
\end{tabular}
\caption{
Pair loading factor $Z_\pm(\xi)$ and pre-acceleration Lorentz factor 
$\gpre(\xi)$ in the prompt radiation front propagating in the external medium 
(with $\mu_e=2$) ahead of the blast wave. 
The radiation spectrum is assumed to be a broken power-law with 
the low-energy photon index $-1$ and the 
high-energy photon index $-2.5$  (these indices are typical for observed GRBs);
the spectrum is assumed to extend to 100~MeV.
\textit{Left:} results for the radiation spectrum with $\Ep=3$~MeV; the exact 
calculation (solid curves) is compared with the cold approximation (dotted curves).
\textit{Right:} results for the radiation spectra with 
$\Ep=100$~keV (dotted), 300~keV (short dashed), 1 MeV (long dashed),
3~MeV (solid), and 10~MeV (dash-dotted).
Note that pair loading is very high ($Z_\pm \sim 10^4$) at $\gpre \sim 10$
where the peak of the GeV flash is emitted (Section~2).
}
\label{fig_fronts}
\end{figure*}

Then we relaxed the cold approximation and obtained $Z_\pm(\xi)$ and 
$\gamma(\xi)$ for bursts with high $\Ep$. \Fig~\ref{fig_fronts} 
shows sample models with $\Ep=1, 3, 10$~MeV, 
$\alpha_1=0$ (photon index $-1$), and $\alpha_2=1.5$ (photon index $-2.5$).
The obtained $Z_\pm(\xi)$ and $\gpre(\xi)$ do not depend on $L_E^{\rm pk}$. 

For comparison, \Fig~\ref{fig_fronts}  (left panel) also shows
the results obtained with the cold approximation,
which are significantly different. 
MeV radiation scattered by the cold plasma is preferentially directed along 
radius (a Klein-Nishina effect),
which reduces the efficiency of pair creation.
One can see that relaxing the cold approximation leads to significantly higher 
$Z_\pm$, mainly because the hot plasma scatters photons through larger 
angles with respect to the
primary collimated beam. The thermal Lorentz factor
of the $e^\pm$ plasma in the radiation front reaches $\gth\approx 3$ in the 
``non-relativistic'' zone where $\gpre\approx 1$; $\gth$ is 
reduced at larger $\xi$ where $\gpre\gg 1$.

%%%%%%%%%%%%%%%%%%%%%%%%%%%%
%%%%%%%%%%%%%%%%%%%%%%%%%%%%

\subsection{Forward shock}

The forward shock propagates in the pair-rich, pre-accelerated medium 
which
is moving with $\gamma<\Gbw$. The shock thermalizes the relative Lorentz factor,
\beq
\label{eq:Grel}
  \Grel=\Gbw\gpre(1-\bbw\bpre)
     \approx \frac{\Gbw}{\gpre(1+\bpre)},
\eeq
where $\bbw=(1-\Gbw^{-2})^{1/2}$ and $\bpre=(1-\gpre^{-2})^{1/2}$.
If there is no energy exchange between $e^\pm$ and ions,
all shocked particles acquire the thermal Lorentz factor 
$\ginj\sim\Grel$ (assuming ``cold'' plasma ahead of the shock, $\gth\sim 1$).
Some energy exchange is, however, expected. Let $\varepsilon_e\leq 1$ be the
fraction of ion energy that is immediately shared with $e^\pm$ due to collective
processes in the shock. 
Then the thermal Lorentz factor of shocked $e^\pm$ is given by
\beq
\label{eq:ginj}
    \ginj=\Grel\left(\gth+\varepsilon_e\,\frac{\mu_e m_p}{Z_\pm m_e}\right),
\eeq
where $\mu_e=1$ for hydrogen and $\mu_e=2$ for heavier ions.
The preheating by the prompt radiation gives $\gth$ comparable to unity (\Sect~3.1);
in \Sect~8 we will discuss an extension of the model that can give $\gth\gg 1$.

In the region of extremely strong pair loading, $Z_\pm\gg 10^3$,
the second term on the right-hand side of \Eq~(\ref{eq:ginj}) is small
compared with the first term, i.e. ions are energetically unimportant.
In this zone, the shock emission is produced by pairs with $\ge\sim\Grel$
regardless of the value of $\varepsilon_e$; the $e^\pm$ pairs dominate the
postshock energy density and quickly radiate this energy away,
leading to nearly 100\% radiative efficiency.

The parameter $\varepsilon_e$ can become important where $Z_\pm\ll 10^4$.
Numerical simulations of electron-ion shocks without pairs show 
$\epse\sim 0.1-0.3$ \citep{sironi_2011}. To our knowledge, there exist 
no calculations of $\varepsilon_e$ for pair-loaded electron-ion shocks;
it is possible that $\epse$ depends on $Z_\pm$. 

The shock may also accelerate a small fraction of electrons/positrons to
Lorentz factors much larger than $\ginj$, forming a nonthermal electron
population.
We assume that most of the shock energy is given to the quasi-thermal
$e^\pm$-ion plasma, and neglect nonthermal particles.
As will be seen below, they are not needed to produce the GeV flash,
and are not expected to dominate the flash energy output.

\subsection{Blast-wave dynamics}

The Lorentz factor $\Gbw$ of the blast wave propagating in the pre-accelerated
medium with a given Lorentz factor $\gpre(R)$ is calculated similarly to
the standard model where the external medium is at rest.
We are particularly interested in the early stage, before the reverse
shock crosses the main part of the ejecta that carries most of the explosion energy.
An estimate for $\Gbw$ at this stage was given in \Sect~2.3. 

In our simulations we use a rather crude model for the blast-wave dynamics. 
Our approach is similar to the ``mechanical'' model of \citet{beloborodov_2006}, 
where the blast-wave material is described by a single Lorentz
factor $\Gbw$, and its evolution with time is
derived from energy and momentum conservation. The pre-acceleration
of the external medium by radiation reduces the pressure in the blast wave.
The blast wave develops where $\gpre<\Gej$, closing the gap between the
radiatively pre-accelerated external medium and the ejecta (B02).
When the reverse shock becomes relativistic ($\Gbw\ll\Gej$) the value of 
$\Gej$ becomes unimportant --- it has no influence on $\Gbw$;
this fact is also seen in the estimate~(\ref{eq:Gbw}).

The relativistic reverse shock crosses the ejecta on an observed timescale
comparable to $\Tb$. At later times the energy supply to the blast wave
from the ejecta drops, and the explosion dynamics switches to the self-similar
regime; we follow this transition in our simulation.
The self-similar blast wave in a wind medium with a low radiative efficiency has
$\Gbw\propto R^{-1/2}$, and with a high radiative efficiency $\Gbw\propto R^{-1}$ \citep{blandford_1976}.

As discussed above, radiative efficiency is close to 100\% during 
the peak of the GeV flash; it can also be high at later phases of the flash 
(see \Sect~8 below). The dynamics of radiative blast waves involves subtle effects.
The large energy losses of the post-shock plasma
imply its quick and significant compression.
In this regime, the forward shock has the Lorentz factor $\Gsh\approx\Gbw$.
There is a thin
shell of fluid immediately behind the shock with Lorentz
factor $2^{-1/2}\Gsh$ (as required by the jump conditions), so the true
profile of the fluid Lorentz factor behind the shock 
is not flat --- there must be a steep change from $2^{-1/2}\Gbw$ to $\Gbw$.
The corresponding velocity profile is consistent with quick compression
of the post-shock plasma --- the expected result of strong radiative losses.
The characteristic thickness of the compression layer behind the shock
is set by the cooling length.

In the radiatively inefficient regime, the blast wave becomes nearly
adiabatic and $\Gsh\approx 2^{1/2}\Gbw$, i.e. the shock runs significantly
faster, leaving more space for the post-shock material. Then the
profile of the fluid Lorentz factor behind the shock is smooth and flat.

We model the transition between the radiative and adiabatic regimes
in a crude way, switching from $\Gsh=\Gbw$ to $\Gsh=2^{1/2}\Gbw$
when radiative efficiency drops below $1/2$.
Full hydrodynamical simulations will be needed in future accurate models.

%############################################################

\section{Radiative transfer}

As long as the GeV flash is dominated by IC scattering of the prompt 
radiation streaming through the blast wave, its light curve can be obtained 
by solving radiative transfer for the prompt photons. The results will describe
the main phase of the flash --- its peak and early decay.
Observations of GeV flashes by {\it Fermi} LAT are typically limited to this 
early phase; e.g. in GRB~080916C it lasts until $\tobs\sim 400$~s (see below).

\begin{figure}[t]
\begin{tabular}{c}
\includegraphics[width=0.45\textwidth]{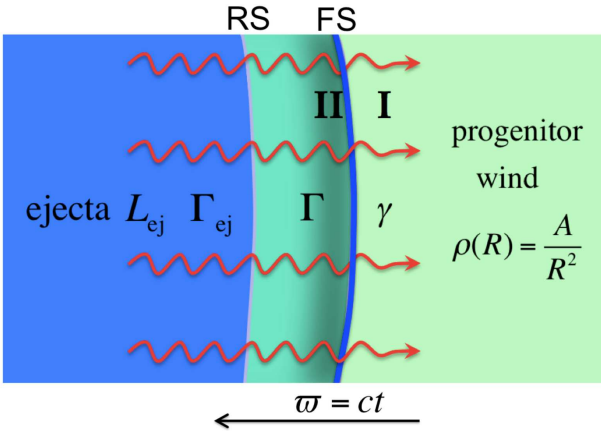}
\end{tabular}
\caption{
Schematic illustration of the transfer problem. Red arrows show 
the prompt MeV radiation streaming from the ejecta and gradually overtaking
the forward shock (FS). The prompt photons can be scattered in the external 
medium ahead of the shock (zone I) or in the shock-heated plasma (zone II).
The coordinate $\x$ measures the distance from the leading edge of the 
radiation front; the unscattered prompt radiation arrives to the observer at time
$\tobs=(1+z)\x/c$. The scattered photons arrive with a delay.
}
\label{fig_scheme}
\end{figure}

Pair loading described in \Sect~3.1 
can also be thought of as a result of radiative transfer of the prompt photons,
but scattered in the external medium {\it ahead} of the blast wave.
One can think of 
both pair loading and flash emission as two parts of one global
transfer problem for the prompt photons (\Fig~\ref{fig_scheme}).
To find an approximate solution to this problem, we divided it into
two zones: ahead of the forward shock (zone~I) and behind the shock (zone~II).
Scattering in zone~I controls the pair loading of the blast wave 
(as it generates MeV photons with large angles).
The GeV flash is mainly produced by scattering in the shock-heated zone~II.

%%%%%%%%%%%%%%%%%%%%%%%%%%%%
%%%%%%%%%%%%%%%%%%%%%%%%%%%%

The result of transfer in zone~I was described in \Sect~3.1.
The solution depends on the prompt radiation spectrum and should be
obtained individually for a given GRB. For a given 
spectral shape (i.e. given $\alpha_1$, $\alpha_2$, $\Ep$)
the obtained $Z_\pm$ and $\gamma$ at the forward shock are functions of the 
GRB energy ahead of the shock,\footnote{
      Our simulation for GRB~080916C also takes into account that
      $\alpha_1$, $\alpha_2$, and $\Ep$ vary during the prompt emission, 
      which affects the relation between $\Eb$ and $\xi$ and the dependence of $\xi$
      on $R$.}
\beq
\label{eq:EFS}
        \Eb=\int_0^{\tFS} \Lb(t)\, dt,
\eeq
where $t=(1+z)^{-1}\tobs$ and $\tFS$ is defined in \Eq~(\ref{eq:xFS}) below.
$\Eb$ determines the value of parameter $\xi$ (see \Eq~(\ref{eq:xi}))
and thus determines $Z_\pm$ and $\gpre$. Note also that $\gpre$ and $Z_\pm$ 
enter our calculation of the blast-wave dynamics $\Gbw(R)$  
(\Sect~\ref{subsect_gbw}), thus the two calculations are coupled and 
we perform them together, integrating over the history of the blast-wave expansion.

Once we obtain solutions for $\Gbw(R)$, $Z_\pm(R)$, and $\gamma(R)$,
we turn to the calculation of photon scattering behind the shock (zone~II).
The blast wave is optically thin, so only a small fraction of the prompt GRB
photons is involved in the radiative transfer.
In addition, multiple IC scattering is strongly suppressed by the 
Klein-Nishina effect at high energies, so one can 
safely use the single scattering approximation. 
One must, however, follow the transfer of
scattered photons through the radiation field, as many of them have high
energies and can easily convert to $e^\pm$ pairs, even though they
have small angles $\theta\sim \Gbw^{-1}$.
The secondary high-energy pairs are Compton cooled by the prompt radiation,
increasing the multiplicity of IC photons.

Monte-Carlo technique is most suitable for this transfer problem.
As the shock passes distance $dR$ it sweeps up
$dN_\pm=Z_\pm(R)n_p 4\pi R^2 dR$
electrons/positrons, where $n_p(R)$ is the proton number density of the external
medium. The shocked particles are heated to $\ginj$ given by \Eq~(\ref{eq:ginj}).
Effectively, $dN_\pm$ hot particles are injected at the shock radius $\RFS$, 
and we follow their cooling behind the shock, track the produced IC photons,
any secondary products that may result from photon absorption, 
and cooling of the secondary pairs.

Particles and photons can be followed on the space-time diagram using
lab-frame time $\tlab$ and radial position $R$ as coordinates.
Note that $R$ is very close to $c\tlab$ everywhere in the relativistic blast wave
(whose characteristic thickness $R/\Gamma^2\ll R$). Therefore, instead of $\tlab$,
it is convenient to use the coordinate $\varpi$ defined by
\beq
   \varpi=c\tlab-R.
\eeq
Then $\varpi=0$ corresponds to the first GRB photons that will be received
at $\tobs=0$, and $\varpi_{\rm GRB}=(1+z)^{-1}c\Tb$ corresponds to the end
of the prompt GRB, $\tobs=\Tb$ (see \Fig~\ref{fig_scheme}).
As long as a particle has coordinate $\varpi<\varpi_{\rm GRB}$, it is exposed
to the prompt GRB photons and can scatter them.
When coordinates $(R,\varpi)$ are used instead of $(R,\tlab)$, one can assume
that all particles in the blast wave have the same radial position $R$, as the
information about the small differences $\Delta R\sim R/\Gbw^2$ is carried by
the coordinate $\varpi$. The blast-wave evolution is fully described by functions
of $R$, e.g. $Z_\pm(R)$, $\Gamma(R)$, etc. The growing radius of the 
expanding blast wave, $R\approx c\tlab$,
now plays the role of a lab-frame time instead of coordinate $\tlab$.

The coordinate $\varpi$ of the forward shock is given by
\beq
\label{eq:xFS}
   \xFS(R)=c\tFS=\int_0^R \frac{dR^\prime}{2\Gsh^2(R^\prime)}.
\eeq
All shocked particles are advected by the expanding blast wave with Lorentz
factor $\Gbw$, and their positions in the prompt radiation front, $\varpi$,
evolve according to
\beq
   d\x=\frac{dR}{2\Gbw^2}.
\eeq

Next, consider an IC photon scattered at $R_{\rm sc}$, $\xsc$ through an angle
$\theta_{\rm sc}$ (measured in the lab frame). The scattered photon 
 propagates along a straight line and 
its angle relative to the radial direction decreases,
\beq
\label{eq:theta}
   \sin\theta(R)=\frac{R_{\rm sc}}{R}\,\sin\theta_{\rm sc}.
\eeq
The photon coordinate $\varpi(R)$ grows according to
\beq
   d\x=(1-\cos\theta)dR.
\eeq
As the IC photon propagates, we evaluate $\gamma$-$\gamma$
opacity along the ray (see below) and check for absorption.
If the photon escapes, its arrival time is
\beq
   \tobs(R_{\rm sc},\xsc,\theta_{\rm sc})=(1+z)\left[\frac{\xsc}{c}
       +\frac{R_{\rm sc}}{c}(1-\cos\theta_{\rm sc})\right].
\eeq

Every scattered photon is drawn from the prompt GRB radiation,
which is assumed to be perfectly collimated at radii of interest,
even when viewed from the rest frame of the blast wave.
The luminosity $\Lb(\tobs)$ and spectrum of the prompt radiation are known
from observations; in the simulations we approximate the prompt spectrum 
by a broken power law.
One can directly calculate the prompt radiation flux at any $R$ and $\x$,
\beq
\label{eq:F}
   F(R,\x)=\frac{\Lb(\tobs)}{4\pi R^2},  \qquad \tobs=(1+z)\frac{\x}{c}.
\eeq
The photon scattering by an electron with a given Lorentz factor $\ge$ is
simulated using the exact Klein-Nishina cross section and drawing the target
photons from the prompt GRB spectrum.

We assume that collective
plasma effects maintain the isotropy of the electron distribution.
This does not imply that the scattered radiation is isotropic
in the fluid frame.
The scattering rate for an electron moving with velocity ${\mathbf v}$
is proportional to $1-{\mathbf v}\cdot{\mathbf n}$ where ${\mathbf n}$ is 
the unit vector in the radial direction (the photon direction before scattering).
Thus, the electron has a higher probability to scatter a photon when
${\mathbf v}\cdot{\mathbf n}<0$. As a result,
IC radiation from isotropic relativistic electrons is significantly anisotropic.
The scattered photons have a higher probability to carry a negative momentum
in the fluid frame, which creates a ``rocket effect'' that tends to accelerate the
blast wave. This effect is neglected in our dynamical model of the explosion
(and should be included in future, more detailed models). However, the anisotropy
of IC radiation is accurately calculated in our Monte-Carlo simulation as we
follow all scattering events individually. The anisotropy impacts the distribution
of photon arrival times measured by a distant observer, leading to an additional
delay (see also \citealt{toma_2009}).

The IC photons can escape or get absorbed by another photon. The absorption
opacity is discussed in detail in \Sect~6 below. Our Monte-Carlo simulation
includes the opacity provided by the main (unscattered) beam of the prompt 
radiation,
\beq
   \kappa_{\gamma\gamma}(\epsilon,\theta)
    \approx \frac{7}{12(1+\alpha)^{5/3}}\,\frac{\sT}{m_ec^3}\,
                          F_\epsilon(\epsilon_{\rm thr}), 
\eeq
where $\theta$ is the angle of the IC photon, $\epsilon=E/m_ec^2$ is its 
dimensionless energy, and $\alpha=-d\ln F_\epsilon/d\ln\epsilon$ is the 
spectral slope of target radiation evaluated near the threshold 
$\epsilon_{\rm thr}=2\epsilon^{-1}(1-\cos\theta)^{-1}$.
As we follow each IC photon, we calculate the absorption opacity along 
its trajectory and check for absorption.
If the photon gets absorbed at some $\xabs$, we inject two new particles
(an $e^\pm$ pair) sharing the energy of the absorbed photon. 
The absorbed photons indirectly contribute to the observed 
emission as they create secondary $e^\pm$ pairs whose IC emission may escape.

%############################################################

\section{GeV flash}

We have applied our transfer simulation 
to GRB~080916C, one of the first GRBs detected by LAT. 
It is an extremely bright burst, with isotropic energy equivalent 
$\sim 9\times 10^{54}$~erg 
\citep{abdo_2009}. The burst duration is $\Tb\approx 100$~s,
which corresponds to $\approx 20$~s when corrected for cosmological
redshift $z\approx 4.35$. \citet{abdo_2009} fitted the prompt emission of
GRB~080916C by the Band function in five consecutive time bins.
We use the prompt emission described by these fits at $E<100$~MeV
as an input of our transfer simulation. 

%%%%%%%%%%%%%%%%%%%%%%%%%%%%
\begin{figure}[t]
\begin{tabular}{c}
\includegraphics[width=0.45\textwidth]{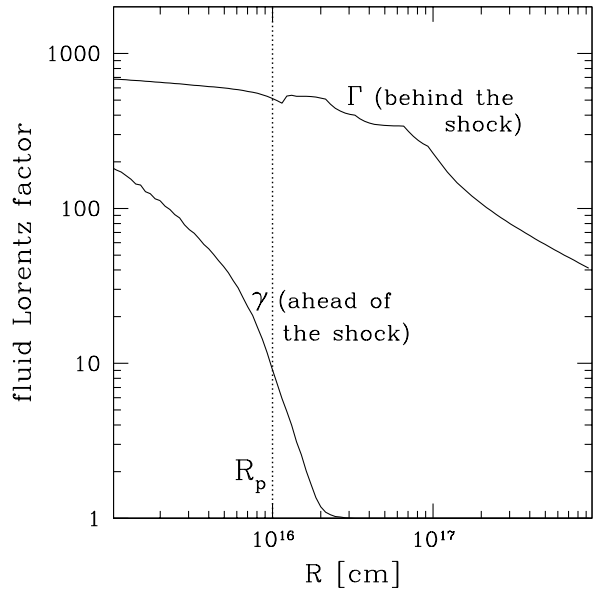}
\end{tabular}
\caption{Lorentz factor of the blast wave ($\Gbw$) and the pre-accelerated 
medium ahead of the blast wave ($\gamma$) in GRB~080916C. 
$R_p$ is the radius where the GeV flash peaks.
The wind density parameter is $A=3\times 10^{11}$~g~cm$^{-1}$.
}
\label{fig_bw}
\end{figure}
%%%%%%%%%%%%%%%%%%%%%%%%%%%%

%%%%%%%%%%%%%%%%%%%%%%%%%%%%
\begin{figure}[h]
\begin{tabular}{c}
\includegraphics[width=0.45\textwidth]{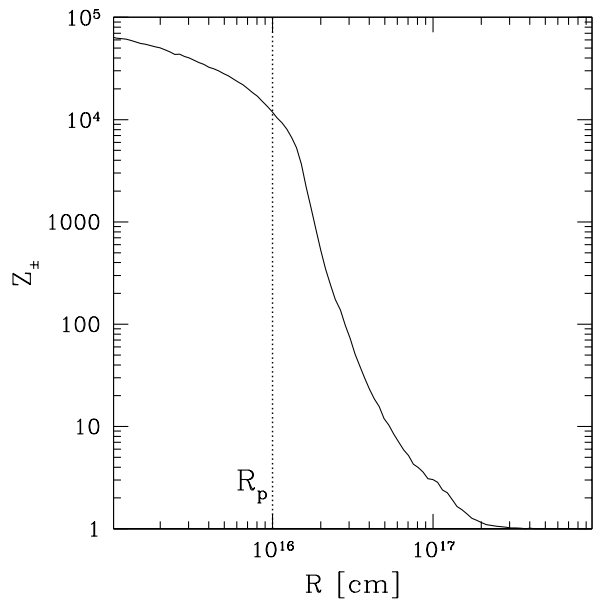}
\end{tabular}
\caption{Pair loading factor of the forward shock.}
\label{fig_Zpm}
\end{figure}
%%%%%%%%%%%%%%%%%%%%%%%%%%%%

%%%%%%%%%%%%%%%%%%%%%%%%%%%%
\begin{figure}[h]
 \begin{tabular}{c}
\includegraphics[width=0.43\textwidth]{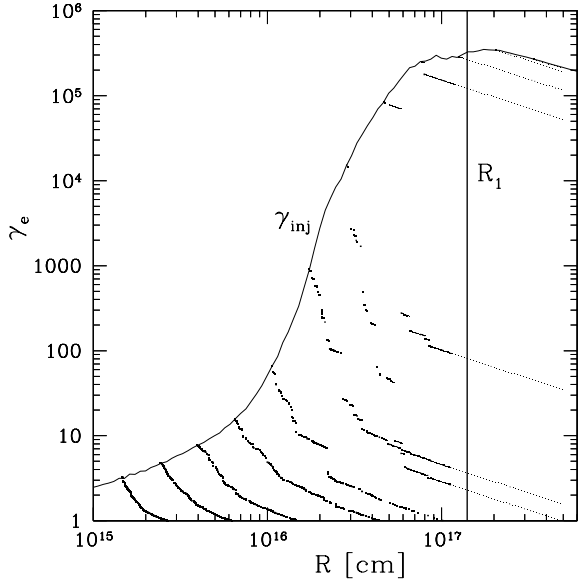}
 \end{tabular}
\caption{Cooling tracks of the shocked particles in the expanding blast wave 
(with $\epse=1$)
exposed to the prompt radiation. Each track starts at the shock front with the 
thermal Lorentz factor $\ge=\ginj$ (shown by the solid curve). 
Compton cooling by the prompt radiation operates (and dominates) at 
radii $R<R_1$; the corresponding tracks are shown by filled squares.
The figure shows one realization of the tracks randomly drawn from our 
Monte-Carlo simulation. Occasional big jumps (the result of large energy loss  
in Klein-Nishina scattering) introduce 
a significant random component, allowing the tracks to cross. At radii $R> R_1$
the prompt radiation decouples from the blast wave 
and no longer can cool it. 
If SSC radiation is neglected, the blast wave becomes adiabatic;
dotted lines show the result of adiabatic cooling.
}
\label{fig_tracks}
\end{figure}
%%%%%%%%%%%%%%%%%%%%%%%%%%%%

The main parameter of the problem is the external density. We consider
the progenitor wind with mass density
\beq
   \rho(R)=\frac{A}{R^2},  \qquad A=\frac{\dot{M}}{4\pi w}.
\eeq
We find that 
$A\approx 3\times 10^{11}$~g~cm$^{-1}$ 
gives a GeV flash consistent with LAT observations,
and therefore in all figures we show the explosion model with this $A$.
The ejecta is assumed to have a high Lorentz factor $\Gej=1200$
and carry energy five times that of the prompt GRB radiation, $\Lej=5\Lb$.
The blast wave is not sensitive to the value of $\Gej$ when $\Gbw\ll\Gej$
(\Sect~\ref{subsect_gbw}).     

Note that the blast wave is optically thin in the region of main interest,
$R\simgt10^{16}$~cm. Its Thomson optical depth at radius $R$ is given by
\beq
   \tau_\pm\approx \frac{Z_\pm \sT A}{\mu_e m_p R}\approx 2\times 10^{-2}
      \left(\frac{Z_\pm}{10^4}\right) A_{11} R_{16}^{-1}.
\eeq
Hereafter we assume $\mu_e=2$ (a progenitor wind that is made of elements 
heavier than hydrogen).

\subsection{Blast wave dynamics, shock heating and cooling}

Figures~\ref{fig_bw} and \ref{fig_Zpm} show the blast-wave dynamics 
$\Gamma(R)$, pair loading $Z_\pm(R)$, and pre-acceleration Lorentz factor 
$\gamma(R)$. The displayed model assumes $\epse=1$; similar results are 
obtained for $\epse=0.1$ and 0. One can see the huge effect of the prompt 
radiation front on the external medium ahead of the blast wave.
The medium is dominated by $e^\pm$ pairs at radii
$R<10^{17}$~cm; $Z_\pm\approx 10^4$ at $10^{16}$~cm.
The prompt radiation accelerates the external medium to a relativistic 
speed at radii $R<2\times 10^{16}$~cm.

The Lorentz factor of the blast wave slowly decreases from 
700 at $R=10^{15}$~cm to 300 at $R\sim 10^{17}$~cm. 
One can notice jumps in the derivative $d\Gbw/dR$.
These jumps are caused by the rough description of the observed prompt radiation
taken from \citet{abdo_2009} 
  --- the burst was divided into five time bins of constant luminosities $\Lb$.
Our simulation assumes $\Lb=0.2\Lej$ (which corresponds to a constant 
radiative efficiency, $\eff=1/6$), and hence the ejecta is discretized into 
five shells with kinetic powers $\Lej=5\Lb$.
The pressure in the reverse shock jumps as it 
crosses the boundary of each shell, which  
affects the blast-wave dynamics.
The reverse shock reaches the end of the ejecta 
at $R\sim 10^{17}$~cm and then the blast wave switches to the self-similar deceleration.
At a comparable radius, Compton 
cooling of the forward shock becomes inefficient (as nearly all prompt radiation
has overtaken the forward shock and decouples from it), 
and the blast wave becomes adiabatic.
In this model, we neglected synchrotron self-Compton (SSC) cooling of the 
blast wave, because for GRB~080916C it becomes important only at late times 
$\tobs>300$~s, where the LAT data ends.

\Fig~\ref{fig_tracks} shows the cooling tracks of the shock-heated particles on the 
$R$-$\ge$ plane. The particles are cooling fast as long as 
the forward shock overlaps with the prompt radiation front, in agreement with 
\Eq~(\ref{eq:tc}). Our simulation assumes that the prompt GRB ends 
at $\x_{\rm GRB}/c=(1+z)^{-1}\Tb\approx 19$~s. The last prompt photons overtake
the forward shock at radius $R_1\approx 1.2\times 10^{17}$~cm, and 
Compton cooling by the prompt radiation ends.

\subsection{Light curve}

\begin{figure}[t]
 \begin{tabular}{c}
\includegraphics[width=0.45\textwidth]{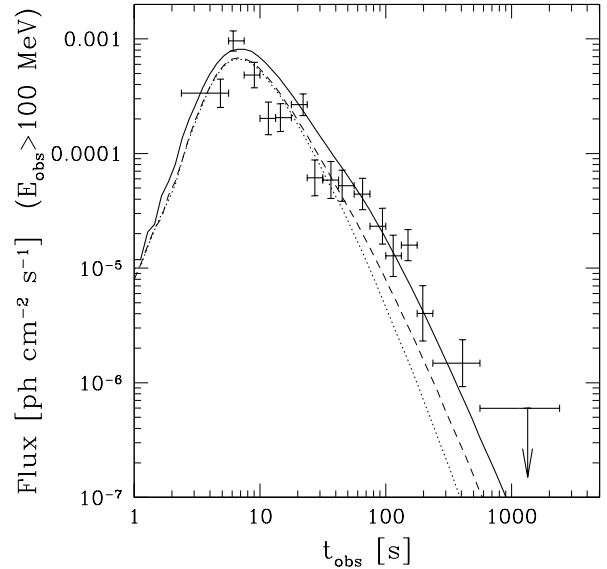}
 \end{tabular}
\caption{Theoretical light curve and data above 100~MeV for GRB~080916C. 
The wind density parameter is $A=3\times 10^{11}$~g~cm$^{-1}$.
To illustrate the effect of $\epse$, we ran the simulation for three cases: 
$\epse=0$ (dotted curve), $0.1$ (dashed curve), and 1 (solid curve).
Data is from \citet{lat_2013}. 
}
\label{fig_lc}
\end{figure}

\Fig~\ref{fig_lc} shows the light curve of high-energy emission 
($E_{\rm obs}>100$~MeV) predicted by the transfer simulation, and 
compares it with the LAT data. 
The peak of the GeV flash at $\tobs\sim 7$~s is dominated by IC emission 
near radius $R_p$ indicated in Figures~\ref{fig_bw} and \ref{fig_Zpm}.

%%%%%%%%%%%%%%%%%%%%%%%%%%%%
%%%%%%%%%%%%%%%%%%%%%%%%%%%%

The shock wave is a weak producer of GeV emission at radii $R<R_p$
because the shock is weak --- it propagates in the medium 
pre-accelerated by the prompt
radiation pressure to a large Lorentz factor $\gamma$, which reduces
the ram pressure in the shock and the thermal Lorentz factor of shocked particles
$\ginj$ (\Eq~\ref{eq:ginj}).
The IC emission of the forward shock appears in the GeV band when $\gamma$ 
decreases to $\sim 10$ and $\ginj$ reaches $\sim 50$.
This condition determines the radius $R_p$ where the GeV flash peaks.
As the shock expands to larger radii $R>R_p$, 
$\ginj$ becomes much greater than 50 and the multiplicity 
of GeV photons saturates at $\mathcal{M} \simlt 10$ (see \Sect~2.1).
Then the decrease of the pair loading factor $Z_\pm$ 
(\Fig~\ref{fig_Zpm})\footnote{The smaller $Z_\pm$ ahead of the shock is partially 
     compensated by the production of secondary particles in the $e^\pm$ cascade 
     behind the shock, which results from $\gamma$-$\gamma$ absorption 
     of high-energy IC photons.}
leads to the decay of the GeV flash.
The decay starts quickly at $R>R_p$, at $\tobs\ll\Tb$,
well before the reverse shock crosses the ejecta, i.e. well before the blast
wave enters the stage of self-similar deceleration. This resolves the puzzle
discussed in \Sect~1.

The production of GeV photons continues as long as the 
shock-heated plasma finds
targets for inverse Compton scattering. Prompt
photons serve as targets until $\xFS=\x_{\rm GRB}$, i.e. until the blast wave 
reaches the
radius $R_1$ where the prompt emission completely overtakes the blast wave,
\beq
\label{eq:R1}
   R_1\approx 2\Gsh^2\,c\,\frac{\Tb}{1+z}.
\eeq
Photons scattered at radius $R_1$ arrive with a significant delay after the last
prompt photons, depending on the scattering angle $\theta$,
\begin{eqnarray}
\nonumber
   \tobs(\theta)&=& \Tb+(1+z)(1-\cos\theta)\frac{R_1}{c} \\
                  & \approx &\Tb\left[1+2\Gsh^2(1-\cos\theta)\right].
\label{eq:tend}
\end{eqnarray}
Here $\Gsh\approx\Gbw$ for a radiative forward shock and 
 $\Gsh^2=2\Gbw^2$ for a shock with a reduced radiative efficiency.
The arrival time given by \Eq~(\ref{eq:tend}) can be much longer than $\Tb$.
For isotropic scattering, the average scattering angle in the fluid frame 
$\tilde{\theta}=\pi/2$ corresponds to $\cos\theta\approx\bbw$ and 
$1-\cos\theta\approx(2\Gbw^2)^{-1}$. This would 
give $\tobs\approx 3\Tb$ 
if the shock is radiatively inefficient at $R_1$, and $\tobs\approx 2\Tb$ 
if it is efficient. In fact, even when the hot electrons are isotropic in the fluid
frame, the scattering is anisotropic ---
the probability of ``backward'' scattering ($\tilde{\theta}>\pi/2$) is larger
than the probability of ``forward'' scattering ($\tilde{\theta}>\pi/2$), as
the backward-moving relativistic electron scatters the collimated prompt
photons with a higher rate. Thomson scattering would give a simple
probability distribution $P(\cos\tilde{\theta})=(1-\cos\tilde{\theta})/2$.
Klein-Nishina corrections change this distribution, however
it remains biased to large $\tilde{\theta}$, delaying the average
arrival time of scattered photons. As a result, a change in the GeV
light curve associated with the end of the target prompt radiation at $R_1$
may be expected at observer time
\beq
    t_1\sim (3-4)\,\Tb.
\eeq

The scattering regime significantly changes over the course of the flash.
The peak at $\tobs\sim\Tp$ is emitted in approximately Thomson regime. 
Indeed, at $R_p$ the shock wave heats the $e^\pm$ pairs to $\ginj\sim 50$ 
while the target radiation density in the fluid frame peaks at 
$\Ep^\prime\sim (2\Gbw)^{-1}\Ep\sim 2$~keV; 
one can see that $\ginj \Ep^\prime/m_ec^2<1$ and hence 
the Klein-Nishina corrections to the scattering cross section are moderate. 
At larger radii (and later observed times) $\ginj$ grows by a few orders of 
magnitude, and the scattering of photons with $\Et\sim\Ep$ is suppressed
by the Klein-Nishina effects. Then the shock wave is mainly cooled by
softer photons of energy 
\beq
  \label{eq:EKN}
\Et\simlt\EKN\sim \frac{\Gbw}{\ginj}\,m_ec^2,
\eeq
and cooling occurs in a regime that is intermediate between the 
Thomson and Klein-Nishina limits.
In this regime, significant luminosity is given to IC photons with 
energies $\EIC$ comparable to the electron energy, and hence the typical 
$\EIC$ weakly depends on the target radiation spectrum. As a result,
the light curve shown in \Fig~\ref{fig_lc} at $\tobs>\Tp$ is not very
sensitive to the spectrum of radiation that provides targets for IC scattering
(we verified this by varying the 
target radiation in our transfer simulation). 
The remaining important condition 
is that the electrons have enough time
to radiate their energy, i.e. cooling is faster 
than the expansion of the blast wave. This condition is satisfied 
(see \Sect~2.2 and \Fig~\ref{fig_tracks}).

The hot electrons see a significant scattering optical depth in 
the target photons of energies $\Et\sim\EKN$. 
Note that the same photons are near the threshold for 
$\gamma$-$\gamma$ reaction with the IC photons of energy 
$\EIC\sim \Gbw\ge m_ec^2$. This implies that the IC photons
see an interesting optical depth to $\gamma$-$\gamma$ absorption
(the $\gamma$-$\gamma$ cross section 
$\sigma_{\gamma\gamma}\simgt 0.1\sT$ is comparable to Compton 
cross section).
In our simulation, we observed significant absorption of IC photons 
and emission from secondary pairs at $\tobs>\Tp$, which has a modest 
impact on
the light curve in \Fig~\ref{fig_lc}. 
It more significantly affects the emission at energies 
$E\gg 1$~GeV (see below).

\subsection{Spectrum}

\Fig~\ref{fig_sp} shows the spectrum of high-energy emission predicted by 
the transfer simulation at $\tobs\sim 2$, 8, and 70~s.
The spectrum is shaped by fast Compton cooling of the shock-heated
$e^\pm$, partial absorption of IC photons by photon-photon collisions,
$\gamma+\gamma\rightarrow e^++e^-$, and cooling of the secondary pairs. 
The spectrum received near the peak of the flash ($\tobs\sim 8$~s) 
is quite flat in the GeV band,
$EL_E\sim \mathrm{const}$, mainly because of the fast evolution of $\ginj$ with 
radius, which implies a quick growth of the maximum IC photon energy 
from $\simlt 1$~GeV to $\simgt 100$~GeV.
As the blast wave expands by a factor of 2 around $R_p\approx 10^{16}$~cm,
$\ginj$ changes by a factor of $\sim 30$ (see \Fig~\ref{fig_tracks}). 
Photons scattered in this region have a broad and flat energy distribution in 
the GeV band, and arrive at comparable times $\tobs$ (which 
vary with the photon angles).

%%%%%%%%%%%%%%%%%%%%%%%%%%%%
\begin{figure}[t]
\begin{tabular}{c}
\includegraphics[width=0.45\textwidth]{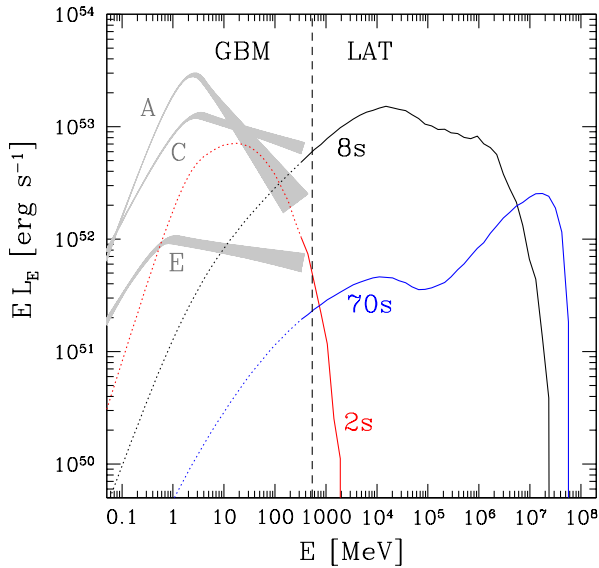}
\end{tabular}
\caption{Spectrum of GRB~080916C in three time windows around
$\tobs\sim 2$, 8, and 70~s from the transfer simulation with $\epse=1$.
Photon energy has been corrected for cosmological redshift $z=4.35$,
$E=(1+z)E_{\rm obs}$.
Vertical dashed line shows the lower boundary of LAT data $(1+z)\times 100$~MeV.
The high energy part (above 100~MeV) shows the 
IC emission from the blast wave, the result of our transfer calculations. 
The gray strips show the GBM data in three time bins A, C, E that roughly  
correspond to $\tobs\sim 2$, 8, and 70~s. The width of the strips indicates the
1-$\sigma$ uncertainty of the spectral fits by the Band function (Sylvain Guiriec, 
private communication).
}
\label{fig_sp}
\end{figure}
%%%%%%%%%%%%%%%%%%%%%%%%%%%%

After the peak, $\tobs>\Tp$, a large fraction of the blast-wave power is 
emitted at energies $E\simgt 100$~GeV.
Absorption is significant for photons with energies $E>10$~GeV; 
however, it never completely suppresses the high-energy emission. This
is an interesting feature of radiative transfer through the pair-loaded blast wave. 
It is related to the fact that the flash peaks when the radiation front has a 
well defined value of 
$\xi\sim 300$ (see \Sect~5.4) and $\xi$ gradually decreases after the peak.
The parameter $\xi$ is a measure of the column density of prompt photons,
and its preferred value $\xi\sim 300$ corresponds to a preferred value of the optical depth to 
$\gamma$-$\gamma$ absorption, $\tau_{\gamma\gamma}$, which 
turns out to be comparable to unity.
The opacity seen by the high-energy IC photons
is dominated by the unscattered, beamed prompt radiation with photon index 
close to $-1$ (energy index $\alpha_1\approx 0$). The resulting optical depth 
is roughly constant at $E\gg 10$~GeV, and its dependence on the emission 
angle $\theta$ is given by 
\beq
   \taugg(x)\approx 0.06 \, x^2 \epsp^{-1}\,\xi,  \qquad 
\eeq
where $x=\theta\Gbw\sim 1$ and $\epsp=\Ep/m_ec^2\sim 10$ in GRB~080916C. 
We used \Eq~(\ref{eq:taugg1}) derived in \Sect~6 below and substituted 
$\alpha=\alpha_1=0$ and $\alpha_2=1.5$. 
A significant fraction of the high-energy photons are emitted within the 
``escape cone''  $\theta\simlt x_{\rm esc} /\Gamma$ where $\taugg\simlt 1$. 

Photons that do not escape  
produce an additional component of ``reprocessed'' high-energy emission
from the secondary pairs. This component creates the flat ``knee'' in the 
spectrum at 1-100~GeV at $\tobs\sim 10-10^2$~s (\Fig~\ref{fig_sp}) and 
leads to the overall two-hump appearance of the high-energy spectrum.

The high-energy spectrum in \Fig~\ref{fig_sp} cuts off at energy $E_{\max}$ 
which increases with time and 
reaches the TeV band at $\tobs\sim 1$~min.
The cutoff is the result of our assumption that 
only thermal heating occurs in the shock wave. 
The Lorentz factor of thermal particles (given by  \Eq~(\ref{eq:ginj})) reaches
$\ginj\sim 10^5$ at late stages of the flash when $Z_\pm$ is reduced.
The thermal particles produce IC photons 
of maximum energy 
$E_{\max}\sim \Gbw\ginj m_ec^2\simlt 10$~TeV.
Emission above $E_{\max}$ is possible 
if the post-shock plasma contains a nonthermal component accelerated at the shock; it would not,
however, make a large contribution to the flash energy and would not
significantly change the GeV emission observed by LAT.

\Fig~\ref{fig_sp} also shows the prompt emission observed 
by Gamma-ray Burst Monitor (GBM) below 100~MeV. Recent analysis of the 
GBM and LAT data shows clear evidence for two separate spectral components
that dominate below and above 100~MeV (\citealt{lat_2013}, Guiriec et al. in 
preparation). This agrees with the theoretical expectation that the prompt 
MeV emission comes from a separate (internal) source at small radii.
Note that its spectrum 
may extend to high energies and contribute to the flux detected
by LAT, mixing with the IC emission from the external shock wave. 
However, the external shock is the stronger source in the GeV band,
especially at late times when the prompt emission declines.

As seen in \Fig~\ref{fig_sp}, the predicted GeV emission from the pair-loaded
external shock starts very soft and quickly hardens as the flash reaches
its peak. The average spectral slope between 
$(1+z)\times 100$~MeV and $(1+z)\times 1$~GeV is consistent with the 
photon index $\sim -2$ observed by LAT \citep{lat_2013}.
Note also that the possible mixing with the (softer) prompt component 
extending to the GeV band can somewhat soften the observed  
spectrum near 1~GeV.

\begin{figure}[t]
\begin{tabular}{c}
\includegraphics[width=0.58\textwidth]{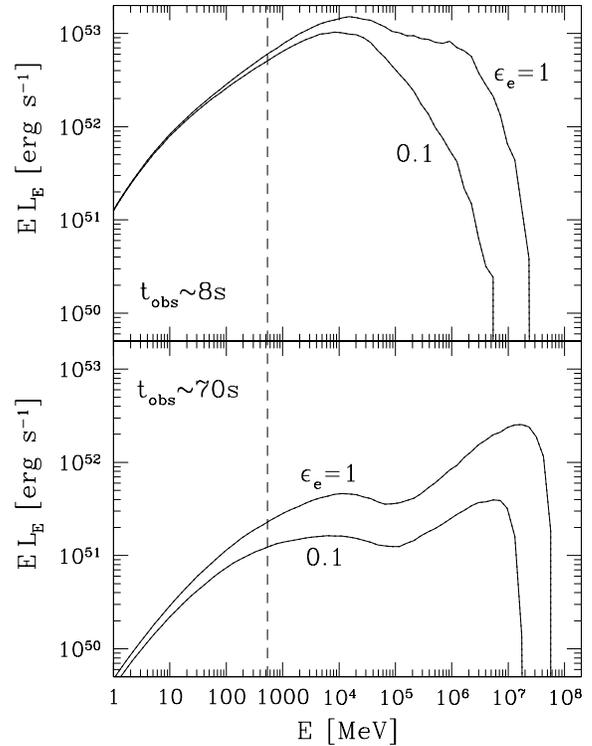}
\end{tabular}
\caption{
Flash spectrum at $\tobs\sim 8$ (upper panel) and $\tobs\sim 70$~s 
(lower panel) for $\epse=1$ and 0.1. Solid curves show the model with no 
magnetic field ($\epsB=0$) and dotted curves show the model with $\epsB=10^{-5}$. 
Photon energy has been corrected for cosmological redshift $z=4.35$,
$E=(1+z)E_{\rm obs}$.
Vertical dashed line shows the lower boundary of LAT data $(1+z)\times 100$~MeV.
}
\label{fig_sp2}
\end{figure}

\Fig~\ref{fig_sp2} compares the predicted high-energy spectrum for $\epse=1$ 
and $\epse=0.1$. The value of $\epse$ makes a significant 
difference for the spectrum at high energies $E\gg 1$~GeV, as the higher 
$\epse$ implies a higher $\ginj$.

%%%%%%%%%%%%%%%%%%%%%%%%%%%%
%%%%%%%%%%%%%%%%%%%%%%%%%%%%

 \subsection{Analytical estimates for $R_p$, $\Gamma$, and $A$}

The radius and Lorentz factor of the blast wave can be quickly 
reconstructed from the observed GeV flash using the following estimates.
The estimates approximately agree with our numerical results for GRB~080916C,
show how the results depend on the GRB parameters, and 
may be applied to other GRBs  with a detected GeV flash.

Let us neglect the variability of the MeV prompt radiation; 
then the shock wave is exposed to radiation of constant luminosity $\Lb$ 
and constant spectrum. The radiation front ahead of the forward shock has 
the energy given by \Eq~(\ref{eq:EFS}), which may be written as
\beq
  \Eb=\Lb\,\tFS.
\eeq  
Here
\beq
\tFS=\frac{\xFS}{c} \approx \frac{R}{2\Gbw^2 c}
\eeq
is the time coordinate of the forward shock,
which is related to the arrival time of the GeV photons by
\beq
\tobs\sim(1+z)\frac{R}{\Gbw^2 c} \approx 2 \, (1+z)\, \tFS.
\label{eq:tobs2}
\eeq
The main parameter $\xi$ that governs pair loading and 
pre-acceleration of the external medium (\Eq~(\ref{eq:xi})) is
\beq
   \xi\approx 650\, L_{54\,} \tFS\, R_{16}^{-2}
   \approx 570 \,  L_{54\,} \left(\frac{\tobs}{1+z}\right)^{-1} \, \left(\frac{\Gamma}{500} \right)^{-4},
   \label{eq:xi2}
\eeq
where $L_{54}=\Lb/10^{54}$~erg~s$^{-1}$. 

The value of $\xi$ at the peak of the flash can be estimated using the approximate relation
(see B02 and \Fig~\ref{fig_fronts}),

\beq
   \gpre\approx \left(\frac{\xi}{\xiacc}\right)^3,  \qquad  \xiacc\approx 100-200,
\eeq
valid in the region of main interest, $1<\xi/\xiacc<3$.
Inverse Compton emission from the shocked electrons peaks at 
$\EIC\sim 1$~GeV 
when $\ginj\sim 2(\EIC/\Ep)^{1/2}$ (\Sect~2), which corresponds to 
\beq
\label{eq:IC}
  \frac{\Gbw}{\gpre}\sim 50,
\eeq
yielding
\beq
\xi 
   \approx 
2  \, \xiacc \, \left( \frac{\Gamma}{500} \right)^{1/3}.
\label{eq:xi3}
\eeq
Combining Equations~(\ref{eq:xi2}) and (\ref{eq:xi3}), we obtain
the radius and Lorentz factor of the blast wave when it emits the peak 
of the GeV flash ($\tobs=T_p$),
\beq
\label{eq:Rp}
   R_p\approx 10^{16}\,
   L_{54}^{6/13} \left(\frac{T_p}{(1+z)\rm ~s}\right)^{7/13} {\rm ~cm},
\eeq
\beq
\Gbw(R_p) \approx 500\,L_{54}^{3/13}  \left(\frac{T_p}{(1+z)\rm ~s}\right)^{-3/13},
\eeq
where $T_p$ is the observed arrival time of the peak.

Using the obtained $\Gamma$ and \Eq~(\ref{eq:Gbw}) one can estimate 
the parameter $A=\dot{M}/4\pi w$ of the wind medium,
\begin{eqnarray}
\nonumber
      A &\approx& \frac{\Lej\, \gpre}{8\pi c^3\Gbw^4} 
        \left(1+\frac{Z_\pm m_e}{\mu_e m_p}\right)^{-1} \\
         &\approx& 10^{11}\,\frac{1-\eff}{\eff}\,
                   L_{54}^{4/13}  \left(\frac{T_p}{(1+z)\rm ~s}\right)^{9/13} {\rm g~cm}^{-1}.
\end{eqnarray}
These estimates assumed that the reverse shock is ultra-relativistic 
($\Gej \gg \Gbw$); it is straightforward to obtain a more general estimate of $A$
using \Eq~(\ref{eq:Gbw_gen}) instead of \Eq~(\ref{eq:Gbw}).

%############################################################

\section{Photon-photon absorption}

The target photons providing opacity for the GeV flash
can be divided into two categories:
(1) the almost perfectly collimated prompt radiation (\Sect~6.1), and 
(2) scattered prompt photons (\Sects~6.2 and 6.3).
The density of scattered radiation is relatively small --- the external 
medium and the blast wave are optically thin even after $e^\pm$ loading, --- 
however, it may provide an interesting contribution to the $\gamma$-$\gamma$ 
opacity, because the scattered photons have larger angles and higher energies.

\subsection{Unscattered prompt radiation}

Let us first evaluate the $\gamma$-$\gamma$ opacity provided by the 
unscattered prompt radiation, which we assume to be perfectly collimated 
at radii where the GeV flash is produced. The absorption optical depth 
seen by a high-energy photon of dimensionless energy $\epsilon=E/\me c^2$ 
propagating at some angle $\theta$ along its path $s$ is given by
\begin{align}
    \taugg(\epsilon) = \iint \frac{F_\epsilon(\epss)}{\me c^3}   \,
     \sigmagg(\epscm) \, (1-\mu) \, d\ln\epss \, ds,
\label{eq:pprod:tau}
\end{align}
where $\sigmagg$ is the cross section for reaction 
$\gamma+\gamma\rightarrow e^++e^-$ in the
center-of-momentum frame of the two colliding photons, $\epscm$ is the
photon energy in this frame, and $\mu=\cos\theta$ describes the angle
between the two photons in the lab frame.
The spectral flux of the target photons is
\begin{align}
       F_\epsilon(\epss)=\frac{L_\epsilon(\epss)}{4\pi R^2},
\label{eq:pprod:F}
\end{align}
where
\begin{align}
\label{eq:Leps}
      L_\epsilon(\epss)=\Lp\, \left(\frac{\epss}{\epsp}\right)^{-\alpha}
\end{align}
is the spectral luminosity of the prompt radiation
and $\epsp$ is the peak/break energy of the prompt GRB
spectrum. For a broken power-law spectrum with indices $\alpha_1$ 
and $\alpha_2$, $\Lp$ is related to the bolometric luminosity $\Lb$ by 
\beq 
      \Lb= \frac{(\alpha_2-\alpha_1)}{(1-\alpha_1)(\alpha_2-1)}\,\Lp\,\epsp.
\eeq

Using the relation $2\epscm^2 = \epsilon \epss(1-\mu)$ to express $\epss$
in terms of $\epscm$ and evaluating the integral over $\epscm$, one finds
\begin{align}
   \taugg =  \psi \,
    \sigmat \int \frac{\Lp}{4\pi \me c^3 R^2} \, 
                                 \left(\frac{\epstrh}{\epsp}\right)^{-\alpha} \, (1-\mu) \, ds,
\label{eq:pprod:tau2}
\end{align}
where
\begin{align}
    \epstrh=\frac{2}{\epsilon(1-\mu)},
\end{align}
and the numerical factor $\psi(\alpha)$ can be approximated as \citep{svensson_1987},
\begin{align}
    \psi(\alpha)  \approx \frac{7}{12(1+\alpha)^{5/3}},  
\end{align}
which is accurate to within 0.3\% in the range $0<\alpha<6$.
The quantity $\psi\sT$ has the meaning of effective cross section for absorption.
The spectral slope $\alpha=\alpha_1$ if $\epstrh\ll\epsp$ and
$\alpha=\alpha_2$ if $\epstrh>\epsp$.

Consider a high-energy photon generated by IC scattering at radius 
$\Rsc$ with angle $\thsc$ relative to the radial direction. 
As the photon propagates, its angle changes according to \Eq~(\ref{eq:theta}).
This change is related to the path element $ds$ by $ds=-R\,d\theta/\sin\theta$, 
and one can express the integral in \Eq~(\ref{eq:pprod:tau2}) as an integral over 
$0<\theta<\thsc$, which yields (in the small-angle approximation 
$\thsc\ll 1$),
\begin{align}
    \taugg(\epsilon,\thsc) = \frac{\sigmat \Lp}{4\pi \me c^3} 
     \frac{\psi(\alpha)}{2^{2\alpha+1}(2\alpha + 3)} \, 
        \frac{(\epsp \epsilon)^{\alpha} \, \thsc^{2\alpha + 2}}{\Rsc}.
\label{eq:pprod:tau3}
\end{align}
Note that $\taugg\rightarrow 0$ if $\thsc\rightarrow 0$. The condition 
$\taugg<1$ defines an escape cone 
$\thsc<\thesc(\epsilon)$ for IC photons of a given energy $\epsilon$.

It is useful to rewrite \Eq~(\ref{eq:pprod:tau3}) as
\beq
    \taugg = \frac{\sigmat \Lp\,\thsc^2}{8\pi \me c^3\Rsc} 
     \frac{\psi(\alpha)}{(2\alpha + 3)} \,\left(\frac{\epstrh}{\epsp}\right)^{-\alpha},
\eeq
where $\epstrh\approx 4(\epsilon\thsc^2)^{-1}$ is the threshold energy 
evaluated at the emission radius $\Rsc$.
High-energy photons produced by the plasma moving with a bulk Lorentz 
factor $\Gbw$ have the characteristic beaming angle $\thsc\sim\Gbw^{-1}$     
(or somewhat larger, because of the anisotropy effect discussed after
\Eq~(\ref{eq:F})). 
It is convenient to describe the photon angle using the variable $x=\Gbw\thsc$,
which is comparable to unity for a typical IC photon. Then the optical depth
may be written as
\beq
\label{eq:taugg1}
   \taugg \approx  \xi\, x^2\,\frac{\Lp}{\Lb}\,
     \frac{\psi(\alpha)}{(2\alpha + 3)} \,\left(\frac{\epstrh}{\epsp}\right)^{-\alpha}.
\eeq
Here $\xi$ is the main physical parameter of the prompt radiation front
given by \Eq~(\ref{eq:xi}), and we estimated $\Eb$ ahead of the forward 
shock as $\Eb\approx \Lb\tFS$ with $\tFS\approx R/2\Gbw^2 c$. The 
peak of the GeV flash occurs where $\xi\sim 300$ (\Sect~5.4).
       
IC photons of energy $\epsilon<\epsilon_1=4\Gbw^2/\epsp x^2$ 
interact with prompt photons $\epss>\epstrh>\epsp$ and 
$\alpha=\alpha_2$; this gives $\taugg<1$.  
Absorption is significant for IC photons with $\epsilon>\epsilon_1$. 
These photons can interact with the low-energy part of the prompt spectrum
$\epss<\epsp$ where 
$\alpha=\alpha_1$. Note that $\alpha_1\approx 0$ (photon index $-1$)
is typical for GRBs, including GRB~080916C. Then $\taugg$ weakly varies with 
$\epsilon$ for $\epsilon>\epsilon_1$, and its value is close to unity
for $\xi\sim 300$. 
      
For GRBs with $\alpha_1 < 0$,
$\taugg$ is maximum at $\epsilon=\epsilon_1$ and decreases at higher
energies. For GRBs with $\alpha_1>0$, $\taugg$ continues to grow with 
$\epsilon>\epsilon_1$ and becomes well above unity. Then the size of the escape 
cone $\thesc$ decreases as a power-law with $\epsilon$, and so does the fraction
of escaping photons. This implies a steeper spectrum where $\taugg\gg 1$ 
(but not an exponential cutoff).

\subsection{Prompt radiation scattered ahead of the forward shock}

High-energy photons from the forward shock have to pass through
the prompt radiation that has been scattered ahead of the shock by the 
pair-loaded and pre-accelerated ambient medium.
The specific intensity of the scattered radiation can be expressed 
as\footnote{
      The factor $d\epsun/d\epssc$ is missing in \Eq~(4) in B02.} (B02)
\begin{align}
      \frac{\Isc(\epssc,\musc,\varpi)}{ \epssc}= 
     \int_{0}^{\x} \frac{d\x^{\prime}}{1-\musc}
		            \frac{F_\epsilon(\epsun)}{\epsun} \, \frac{Z_\pm n_0}  {2\pi} 
     \frac{d\sigma}{d\musc} \frac{d\epsun}{d\epssc}. 
\label{eq:pprod:Isc}
\end{align}
Here $F_\epsilon$ is the spectral flux of prompt radiation,
$\epsun$ is the prompt photon energy (before scattering),
$\musc=\cos\theta_{\rm sc}$ describes the scattering angle,
and $\epssc$ is the photon energy after scattering; 
$Z_\pm(\x^\prime)$ is the pair loading factor, and $n_0$ is the external 
electron density before $e^\pm$ loading.
The integral is taken over the Lagrangian coordinate $\x=ct-R$ that measures 
the distance inside the prompt radiation front; $d\x/(1-\musc)$ 
is the elementary path length along the scattered photon trajectory in the 
lab frame.

We are interested in the optical depth $\taugg$ created by the
scattered radiation, as seen by a high-energy photon of energy $\epsilon$
emitted by the shock wave.
The photon has an angle $\theta\sim\Gbw^{-1}$, which is much smaller
than the typical angles of the target photons  $\theta_{\rm sc}\sim \gpre^{-1}$
(where $\gpre=(1-\beta^2)^{-1/2}$ is the Lorentz factor of the pair-loaded 
medium accelerated by the radiation front).
Therefore, here the high-energy IC photon 
may be approximated as perfectly collimated in the radial direction, $\theta=0$.
Then,
\begin{align}
   \taugg(\epsilon) = 2\pi R \iint \frac{\Isc(\epssc,\musc)}{\epssc \me c^3} \, 
         \sigmagg(\epscm) (1-\musc) \, d\musc \, d\epssc.
\label{eq:pprod:tausc}
\end{align}
Following B02, we will
make the simplifying assumption that the prompt radiation is scattered at 
$90^{\circ}$ in the local rest frame of the medium (which corresponds to
$\musc=\beta$ in the lab frame), and approximate the Thomson cross-section 
as $d\sigma/d\musc \approx \sigmat \delta(\musc-\beta)$. Then we obtain,
\begin{align}
     \taugg(\epsilon) =  \psi\, \sigmat R \, n_0\,\frac{\Lp}{\Lb}
      \int_{0}^{\xi} d\xi^{\prime} \,  Z_\pm (\xi^{\prime}) \,
       \left(\frac{\epstrh}{\epsp}\right)^{-\alpha},
\label{eq:pprod:tausc2}
\end{align}
where $\epstrh = 2(1+\beta)/\epsilon(1-\beta)$ 
is the pair-production threshold energy for the prompt photon (before scattering)
for interaction with a high-energy photon $\epsilon$, and 
$\xi = \x \sigmat \Lb/4\pi \me c^3 R^2$.

The optical depth given by Equation (\ref{eq:pprod:tausc2}) can 
be understood as follows. The column density of electrons exposed to the 
prompt radiation is $\sim R\, Z_\pm n_0$ (accounting for pair loading).
Each electron at coordinate $\x$ in the radiation front has scattered
approximately $\xi/\epsp$ photons, 
and hence the column density of scattered photons is 
$\sim n_0 R\, Z_\pm(\xi)\,\xi/\epsp$.
A fraction  $(\epsp \Lp/\Lb) (\epstrh/\epsp)^{-\alpha}$ of these photons
are near the threshold for pair production,
where the average $\gamma$-$\gamma$ cross section is large,
$\sigma_{\gamma\gamma}\sim \psi\, \sigmat$.

Consider a simplified analytical model of the radiation front 
in the region where $1<\gamma\lesssim 30$ (B02),
\begin{align}
   \gamma = \left(\frac{\xi}{\xiacc}\right)^3, \qquad \Zpm 
                 = \Zacc \left(\frac{\xi}{\xiacc}\right)^2,
\end{align}
where $\xiacc\approx 100-200$ (the more accurate front structure is shown 
in \Fig~\ref{fig_fronts}). Then one
can evaluate the integral in Equation (\ref{eq:pprod:tausc2}) using
$Z_\pm\, d\xi = \Zacc\xiacc \, d\gamma/3$ and 
$\epstrh \approx 8\gamma^2/\epsilon$, which yields
\begin{align}
    \taugg(\epsilon) = \frac{ \psi(\alpha)\,\Zacc \, \xiacc}
          {3 (1-2\alpha) 2^{3\alpha}} \,  \frac{\Lp}{\Lb}  \, 
          (\epsp \epsilon)^{\alpha} \, \gamma^{1-2\alpha}\, 
     \tau_0,
\label{eq:pprod:tausc3}
\end{align}
where 
$\tau_0=\sigmat R \,n_0(R)$
is the Thomson optical depth through the progenitor wind and $\gamma$ is 
the pre-acceleration Lorentz factor at the location of the forward shock.

The power-law segment of the Band spectrum that provides the dominant 
contribution to $\taugg$ is determined by comparing $\epsp$ and $\epstrh$;
the lower-energy segment dominates if
\begin{align}
    \epsilon > \frac{8\gamma^2}{\epsp}.
\label{eq:pprod:epseps}
\end{align}
Most of the GeV flash is emitted at radii where the pre-acceleration 
Lorentz factor $\gpre\simlt 10$, and the condition (\ref{eq:pprod:epseps}) 
is satisfied for $\epsilon\simgt 10^3$. Then $\alpha=\alpha_1$ in 
\Eq~(\ref{eq:pprod:tausc3}). For the typical $\alpha_1\approx 0$, one finds
 that $\taugg$ at high energies does not depend on $\epsilon$ and its value 
is small, $\taugg<1$, for $\tau_0\sim 5\times 10^{-6}$ expected for the progenitor
wind at the flash radius $R_p\sim 10^{16}$~cm. In particular, for $\alpha_1=0$
and $\alpha_2=1.5$ (typical for GRBs) we obtain $\Lp/\Lb=(3\epsp)^{-1}$ and
\beq
     \taugg\sim 10^4 \epsp^{-1}\,\gpre\,\tau_0<1.
\eeq
    
Our conclusion that the scattered radiation provides a small $\taugg<1$
is different from the estimates in B02 where the radiation scattered in the wind
medium was found to block any GeV emission. There are two reasons for
this difference. First, B02 considered less luminous bursts where 
the pair-loaded region had a smaller radius and hence a larger 
$\tau_0\propto R^{-1}$.
Less luminous bursts also have smaller $\ep$.
Second, the estimates in \Sect~6.3 in B02 confused the photon index with 
the energy index of the prompt GRB spectrum, 
leading to an overestimation of $\taugg$.

\subsection{Prompt radiation scattered behind the forward shock}

The plasma immediately behind the forward shock has an ultra-relativistic 
temperature and here scattering produces high-energy IC photons.
The high-energy photons may interact between themselves. 
An exact calculation of this ``self-absorption'' of 
the GeV flash would require a full nonlinear simulation of radiative transfer.
A simple estimate suggests that the self-absorption effect is not strong 
in GRB~080916C. The isotropic equivalent of the photon number in the flash 
is $N_{\rm GeV}\sim 10^{57}$, and the column density of GeV
photons is $\sim N_{\rm GeV}/4\pi R^2$. This gives an upper bound on 
the absorption optical depth provided by the GeV photons,
\beq
    \taugg\simlt \frac{\sigma_{\gamma\gamma} N_{\rm GeV}}{4\pi R^2}
      \approx 0.1 \left(\frac{N_{\rm GeV}}{10^{57}}\right)\, R_{16}^{-2},
\eeq
where we estimated the effective cross section 
$\sigma_{\gamma\gamma}=\psi\, \sT$ and assumed the spectral index 
$\alpha\sim 1$  (photon index $\sim 2$) in the GeV band, which gives 
$\psi(\alpha)\sim 0.2$.

Further downstream of the shock the plasma cools and accumulates 
in the blast wave. The optical depth of this cold plasma is 
$\tau_\pm= Z_\pm \tau_0$
(pair annihilation is negligible).
It scatters the prompt photons with a moderate change in photon energy
and a typical scattering angle $\theta_{\rm sc}\sim \Gbw^{-1}$.
Some of these photons may overtake the GeV photons emitted immediately
behind the shock and contribute to the absorption opacity seen by the GeV 
photons. Their contribution is
small compared to  $\taugg$ of 
photons scattered ahead of the shock (\Sect~6.2).
The numbers of photons scattered ahead and behind the shock are
comparable, however the angles of photons 
scattered
ahead of the shock are 
much larger, making them more important 
targets for photon-photon absorption.

\subsection{Summary}

The unscatterred collimated prompt radiation dominates the 
$\gamma$-$\gamma$ opacity seen by the GeV photons. 
The corresponding optical depth $\taugg$ is evaluated in \Sect~6.1;
it is shown to be small at energies $E\ll 30$~GeV and comparable 
to unity at higher energies.
Prompt photons scattered ahead or behind the shock provide an 
additional small contribution to $\taugg$, which may be neglected, 
at least for the GeV flash in GRB~080916C.

%############################################################

\section{Synchrotron emission}

The presence of a magnetic field in the blast wave can have 
three
observational effects. (1) If the field is strong, synchrotron losses of the shocked plasma
can compete with its IC cooling by the prompt radiation; this would weaken the 
GeV flash. (2) Synchrotron losses 
give emission in softer bands, e.g. optical or X-rays, providing an additional test for the 
pair-dominated flash mechanism. 
(3) Synchrotron photons may become the main targets for IC scattering by the 
high-energy electrons in the blast wave, which can affect the observed 
light curve and spectrum of high-energy emission.

\subsection{Cooling rate and the characteristic photon energy}

The competition between synchrotron cooling and Compton cooling by the prompt 
radiation was discussed by \citet{beloborodov_2005}. The two contributions to the cooling 
rate of isotropic electrons with a thermal Lorentz factor $\ge\gg 1$ are given by
\beq
   \dot{E}_{\rm syn} =-\frac{4}{3}\sT U_B^\prime c\ge^2,  \qquad
   \dot{E}_{\rm IC} \approx -\frac{4}{3}\sT \UT^\prime c\ge^2,
\eeq
where $U_B^\prime$ is the magnetic energy density, and $\UT^\prime$ is 
the energy density in the prompt photons of energy $E<\EKN$ (\Eq~(\ref{eq:EKN}))
which can be scattered with approximately Thomson cross section; 
$U_B^\prime$ and $\UT^\prime$ are measured in the fluid frame. 
We include only the unscattered prompt radiation in $\UT^\prime$, 
assuming that it  dominates Compton cooling of the blast wave;
the density of synchrotron radiation from the blast wave itself is 
assumed to be relatively small. Then,
\begin{eqnarray}
  \UT^\prime\approx \fT\,U^\prime, \quad \fT \approx \left\{ 
                                         \begin{array}{ll}
       1  &  \EKN\gg\Ep \\
 \displaystyle{ \left( \frac{\EKN}{\Ep} \right)^{-\alpha_1+1}} & \EKN < \Ep 
                                         \end{array}
                                                          \right.
\end{eqnarray}
where 
\beq
    U^\prime=\frac{\Lb}{16\pi c\, R^2 \Gbw^2}
\eeq
is the energy density of the prompt radiation in the fluid frame, and $\alpha_1$ is 
the spectral index of radiation at photon energies $E<\Ep$. 

The magnetic energy density behind the shock may be expressed in the 
standard form using the parameter $\epsB$,
\beq
\label{eq:UB}
   U_B^\prime= 3 \epsB P_f =\epsB\,\frac{4 \rho c^2 \Gbw^2}{\gpre(1+\bpre)},
\eeq
where $\rho$ is the mass density of the external medium and 
$\gpre=(1-\bpre^2)^{-1/2}$ is its pre-acceleration Lorentz factor; 
we neglected the increase in $\rho$ due to $e^\pm$ pairs loaded 
ahead of the shock.
We focus here on the main phase of the GeV flash before the 
reverse shock has crossed the ejecta. Then \Eq~(\ref{eq:Gbw}) may 
be used to obtain another expression for $U_B^\prime$, 
\beq
\label{eq:UB1}
    U_B^\prime\approx\frac{\epsB \Lej}{4\pi\,c\,R^2\Gbw^2}.
\eeq
The ratio of synchrotron and Compton cooling rates is then given by
\beq
\label{eq:ratio}
    \frac{\dot{E}_{\rm syn}}{\dot{E}_{\rm IC}}\approx\frac{U_B^\prime}{\UT^\prime}
    \approx \frac{4\,\epsB\Lej}{\fT \Lb}.
\eeq
The numerical factor $\fT$ is comparable to unity at the peak of the GeV flash,
when the forward shock heats the plasma to $\ginj\sim 10^2$. 
After the peak, $\ginj$ increases, however the flash light curve shown
in \Fig~\ref{fig_lc} is still dominated by particles cooled to $\ge\sim 10^2$, with $\fT\sim 1$.

The characteristic energy of synchrotron photons is given by
\beq
\label{eq:Es}
   E_s\approx 0.2\,\Gbw\ge^2 \,\hbar \frac{eB^\prime}{m_ec},
\eeq
where $B^\prime=(8\pi U_B^\prime)^{1/2}$ is the magnetic field measured 
in the fluid frame. Using \Eq~(\ref{eq:UB1}) one obtains,
\beq
\label{eq:Es1}
  E_s\sim 20\,\epsB^{1/2} \left(\frac{\ge}{100}\right)^2 
              \left(\frac{\Lej}{10^{54}{\rm ~erg~s}^{-1}}\right)^{1/2} R_{16}^{-1} {\rm ~eV}.
\eeq
Most of the synchrotron power is emitted by particles with $\ge\sim\ginj$.
As the blast wave expands from $R\sim 10^{15}$~cm to $10^{17}$~cm,
$\ginj(R)$ evolves from low values $\sim 1$ to $\sim 10^2$ 
(at the peak of the GeV flash) to $\sim 10^4-10^5$, see \Eq~(\ref{eq:ginj})
and Figure~\ref{fig_tracks}. As a result, $E_s(\ginj)$ evolves 
by a huge factor $\sim 10^6$, and hence the blast wave must produce 
broad-band synchrotron radiation. The emitted synchrotron power may be 
estimated using \Eq~(\ref{eq:ratio}) with $\fT$ that corresponds to $\ginj$. 
Moderately high $\epsB\simgt 10^{-5}$ would imply strong synchrotron 
emission in the hard X-ray band. It can easily conflict with the observed 
radiation spectrum, which can be used to infer an upper limit 
$\epsB^{\max}\sim 10^{-5}$ for GRB~080916C.

\subsection{Optical flash}

If one is interested in radiation in a fixed spectral band, e.g. 
optical $E\sim 2\,(1+z)$~eV, the observed emission will be dominated by particles 
that have cooled behind the shock to Lorentz factor $\ge=\go$ such that 
$E_s(\go)\sim 2\,(1+z)$~eV. From \Eq~(\ref{eq:Es1}) one finds 
\begin{equation}
    \go\sim 
    10^3\, (\epsB/10^{-6})^{-1/4}
    L_{\rm ej,54}^{-1/4} R_{16}^{1/2}(1+z)^{1/2}.
\end{equation}
A more accurate expression for $\go$ may be obtained from 
\Eq~(\ref{eq:Es}) using \Eq~(\ref{eq:UB}),
\beq
     \go\approx \frac{10^4}{\Gbw} \left[\frac{\gpre(1+\bpre)}
                      {\epsB\rho c^2}\right]^{1/4} (1+z)^{1/2}.   
\eeq
In the blast wave with pure thermal heating, optical emission 
remains negligible until $\ginj(R)$ exceeds $\go$; the optical light curve is 
expected to reach its peak at this point. This happens soon after the peak of 
the GeV flash.

The subsequent decay of the optical flash can be described using the 
following estimate for the optical luminosity,
\beq
\label{eq:opt1}
    \Lo\sim EL_E\sim \frac{dN_{\pm}}{dt}\,\Gbw\,\frac{\go m_ec^2}{2} \,\fsyn,
\eeq
where $t=(1+z)^{-1}\tobs$, $N_{\pm}$ is the number of electrons/positrons 
cooling behind the shock, and
\beq
   \fsyn=\frac{\dot{E}_{\rm syn}(\go)}{\dot{E}_{\rm IC}(\go)+\dot{E}_{\rm syn}(\go)}
          \approx \frac{U_B^\prime}{\UT^\prime(\go)}.
\eeq 
\Eq~(\ref{eq:opt1}) states that each particle emits in the optical band a fraction 
$\sim\fsyn/2$ of its energy in the lab frame, $\sim \Gbw\,\go m_ec^2$,
as $\ge$ decreases from $\go$ to $\go/2$. 
The emitted energy $\sim \Gbw\go m_ec^2/2$ is shared by IC and synchrotron 
photons; in our case the IC losses dominate and the synchrotron fraction 
$\fsyn\ll 1$ is given by \Eq~(\ref{eq:ratio}). Then we obtain,
\begin{eqnarray}
\nonumber
   \Lo\sim 10^{49} R_{16} Z_\pm \left[\gpre(1+\bpre)\epsB A_{12}\right]^{1/2} \\
     \times \frac{\Lej}{\Lb}\,\epsp\,(1+z) {\rm ~erg~s}^{-1}.
\label{eq:Lopt}
\end{eqnarray}
Here we used $dN_{\pm}/dt\sim Z_\pm(4\pi R^3 \rho/\mu_e m_pt)$ and 
$t\sim R/c\Gbw^2$.
\Eq~(\ref{eq:Lopt}) shows that the decay of the optical flash is controlled by the 
evolution of the factor $Z_\pm R[\gpre(1+\bpre)]^{1/2}$ with time $t$.
This evolution is fast; when approximated by a power law $t^{-a}$ its slope 
is $a\sim -2$.  One can also see from \Eq~(\ref{eq:Lopt}) that the optical flash 
is extremely bright even for a modest $\epsB\sim 10^{-6}$.
Its peak occurs where $Z_\pm\sim 10^2$ and 
can reach an optical luminosity $\Lo\sim 10^{50}$~erg~s$^{-1}$.

In summary, the peak luminosity of the optical flash is achieved when $\ginj$
exceeds $\go$. This typically happens at $\tobs \sim 10(1+z)$~s. The optical flash 
can be extremely bright, but it quickly decays. We find that its luminosity drops by 
a factor of $10^{-2}$ as $\tobs$ grows by a factor of 10, mainly because of the 
decreasing pair loading factor $Z_\pm$. At later times the prompt radiation 
decouples from the blast wave and the Compton cooling ends, which implies 
the end of the fast decay; then the optical flash should evolve to normal optical 
afterglow.

Note that the $e^\pm$ pairs collected at $R\simlt 10^{17}$~cm are Compton cooled 
to a low temperature and do not contribute to the afterglow emission at late times.
This is in contrast to explosion models where the prompt radiation quickly decouples 
from the blast wave and Compton cooling is inefficient; in this case the blast wave
would carry slowly cooling pairs and the synchrotron afterglow would have a long 
``memory'' of pair loading \citep{beloborodov_2005b}.

%############################################################

\section{Impact of the GeV flash on the external medium}

Our transfer simulations described in \Sect~5 show that some of the produced 
high-energy photons do not escape --- they are absorbed by the prompt radiation 
beam and convert to $e^\pm$ pairs. Most of the conversion events occur behind 
the forward shock and join the shocked plasma moving with Lorentz factor $\Gbw$,
however a small fraction convert ahead of the shock and join the external
medium, which moves with a much smaller Lorentz factor $\gpre$.
These rare events create particles of very high energies (GeV-TeV) in the 
external medium, depositing their energy and momentum.
Thus, the GeV flash itself creates additional pre-heating and pre-acceleration 
of the external medium, which was not taken into account in our model of the 
radiation front in \Sect~3.1. We now estimate this effect and its implications.

\subsection{Fraction of the flash power deposited ahead of the shock}

First, let us roughly estimate the fraction of the flash power that converts to 
$e^\pm$ pairs ahead of the shock wave. 
Only photons with sufficiently small angles can overtake the forward shock,
\beq
   \theta<\theta_{\max}=\Gsh^{-1}.
\eeq
For the simplest estimate, we picture the flash source as an infinitesimally 
thin shell behind the shock (the fast-cooling limit) and assume that only 
photons emitted with $\theta<\theta_{\max}$ have a chance to convert ahead 
of the shock. The absorption optical depth $\taugg$ seen by 
these photons is given by \Eq~(\ref{eq:pprod:tau3}); it increases with $\theta$ and 
is maximum at  
$\theta_{\max}$. The deposited power ahead of the shock may be written as
\beq
\label{eq:Lpm}
   \Lpm= \zeta \int  \taugg(\epsilon,\theta_{\max})\,L_\epsilon\,d\epsilon,
\eeq
where 
\begin{eqnarray}
\nonumber
   \taugg(\epsilon,\theta_{\max}) 
      \approx \frac{\psi(\alpha)}{2^{2\alpha+1}(2\alpha + 3)}
        \frac{\sigmat \Lp(\epsp \epsilon)^{\alpha}}{4\pi \me c^3R\,\Gsh^{2\alpha+2}},
\end{eqnarray}
$L_\epsilon$ is the flash spectrum, 
and $\zeta=0.01-0.1$ is a numerical factor determined by the angular 
distribution and spectrum of the flash radiation.
The spectral slope of the target radiation, $\alpha$, is determined as follows.
The main target photons contributing to $\taugg$ have energies
\beq
  \epsilon_t\sim 2\epstrh=\frac{4}{\epsilon(1-\cos\theta)}
     \approx \frac{8\Gsh^2}{\epsilon},
\eeq
which should be compared with $\epsp\sim 10$. This gives,
\begin{eqnarray}
  \alpha=\left\{\begin{array}{ll}
        \alpha_2, & \quad \epsilon<8\,\epsp^{-1}\,\Gsh^2 \\
        \alpha_1, & \quad \epsilon>8\,\epsp^{-1}\,\Gsh^2
                      \end{array}
             \right.
\end{eqnarray}
where the characteristic $\epsilon_1=8\Gsh^2/\epsp$ corresponds to photon 
energy $\epsilon_1 m_ec^2\sim 10^2$~GeV. The flash spectrum extends above 
$\epsilon_1$ after the peak time $\Tp$, when $\ginj$ exceeds $\sim\Gsh$; then 
photons with $\epsilon>\epsilon_1$ make the main contribution to the integral in 
\Eq~(\ref{eq:Lpm}), and $\taugg$ should be evaluated with $\alpha=\alpha_1$. 
In particular, for $\alpha_1=0$ we obtain
\beq
\label{eq:Lpm1}
  \frac{\Lpm}{\Lflash}\sim 0.1\,\zeta\,\frac{\sigmat \Lp}{4\pi m_ec^3 R\,\Gsh^2},
\eeq
where we assumed that a large fraction of the flash luminosity $\Lflash$ is emitted
above $\sim 10^2$~GeV; this assumption is satisfied in the self-consistent model,
as we show below.

\subsection{Pre-heating and pre-acceleration}

The injection of power $L_\pm$ into the external medium can be described
as inelastic collision which heats and accelerates the medium.
Consider an external mass shell 
\beq
   dm=4\pi R^2 \rho\, dR=4\pi A\, dR.
\eeq
It first interacts with the prompt radiation and then it is exposed to the high-energy 
flash photons, which deposit energy,
\beq
   dE_\pm\sim \Lpm\, \frac{dR}{2\Gsh^2c}.
\eeq
This energy is deposited in the form of ultra-relativistic $e^\pm$ pairs, which 
are expected to immediately share their momentum $dE_\pm/c$ with 
the medium through collective processes (B02). 
The GeV flash accelerates the medium to a high Lorentz factor 
$\gamma^\prime\gg 1$ if 
\beq
\label{eq:G}
   G\equiv\frac{dE_\pm}{dmc^2}=\frac{L_\pm}{8\pi c^3 \Gsh^2 A}\gg 1.
\eeq
The deposited energy $dE_\pm$ is shared between the bulk kinetic energy of the 
accelerated medium and its internal energy (i.e. heat).
The ultra-relativistic pairs can scatter the prompt radiation ahead
of the forward shock; however, since the pairs are isotropic in the 
fluid frame, the produced high-energy photons have large angles and quickly 
convert to $e^\pm$ pairs, which join the medium.\footnote{This cascade in the 
    external medium has a moderate effect on pair multiplicity $Z_\pm$. The 
    high-energy particles injected by the flash radiation are relatively close to 
    the forward shock and have time for a moderate number of scatterings before 
    they are swept by the shock. A dedicated numerical simulation will be needed
    to quantify this effect.} 

For simplicity, let us consider radii where the pre-acceleration by the prompt 
radiation is not significant ($R>2\times 10^{16}$~cm, see \Fig~\ref{fig_bw}), so 
that we can isolate the effect of the GeV flash.
We can evaluate the Lorentz factor gained by the shell, $\gamma^\prime$, 
and its new rest-mass $dm^\prime$ (which includes the deposited heat) 
from the energy and momentum conservation laws, 
\begin{eqnarray}
   dm+\frac{dE_\pm}{c^2} & = & \gamma^\prime dm^\prime , \\
          \frac{dE_\pm}{c^2} & = & \gamma^\prime\beta^\prime 
                    dm^\prime.
\end{eqnarray}
This gives,
\beq
    \gh\equiv\frac{dm^\prime}{dm}=\left(2G+1\right)^{1/2},
\eeq
\beq
   \gamma^\prime\beta^\prime=\frac{G}{\gh}.
\eeq
Also note the relation,
\beq
\label{eq:rel}
  \gh=\gamma^\prime(1+\beta^\prime).
\eeq

It is easy to see that $G\gg 1$ is expected, which implies a strong impact of 
the flash on the external medium, $\gh\gg 1$ and $\gamma^\prime\gg 1$.
Indeed, substituting \Eq~(\ref{eq:Lpm1}) into \Eq~(\ref{eq:G}) and using the 
simple estimate for the blast-wave Lorentz factor $\Gbw^4\sim \Lej/16\pi c^3 A$ 
(see \Eq~(\ref{eq:Gbw}) and \Eq~(\ref{eq:Gbw1}) below), one obtains
\beq
   G\sim \frac{0.1\zeta\sT \Lp}{2\pi m_ec^3 R}\,\frac{\Gbw^4}{\Gsh^4}\,
   \frac{\Lflash}{\Lej}
  \sim 0.4 \zeta \xi \frac{\Gbw^4}{\Gsh^2}  \frac{\Lp}{\Lb} \frac{\Lflash}{\Lej}
\eeq
which gives a typical 
$G\sim 10^2-10^3$.
The value of $G$ is strongly 
reduced at smaller radii where the prompt radiation pre-accelerates the 
external medium to $\gpre\gg 1$. The effect of $G\gg 1$ should develop 
soon after the peak of the GeV flash, when $\ginj>10^2$ and $\gamma<10$.

\subsection{Effect on the blast wave Lorentz factor}

We now estimate the effect of pre-acceleration and pre-heating by the 
flash radiation on the blast-wave Lorentz factor $\Gbw$. Similar to \Sect~2.3
we consider sufficiently early times ($\tobs<\Tb$) and use the pressure 
balance between the forward and reverse shock, $P_f\sim P_r$, for a rough 
estimate. On the other hand, to isolate the effect of the flash, we consider late 
enough times when the prompt radiation does not significantly pre-accelerate
the medium, $\gpre\approx 1$. Then \Eq~(\ref{eq:Gbw}), with $\gpre$ replaced 
by $\gamma^\prime$ and $Z_\pm\ll \mu_e m_p/m_e$, gives
\beq
\label{eq:Gbw1}
   \Gbw^4\approx \frac{\Lej}{16\pi c^3 A}.
\eeq
The result is the same as if there were no effect of the flash on the external 
medium --- the terms $\gamma^\prime(1+\beta^\prime)$ and $\gh$ cancel 
(see \Eq~(\ref{eq:rel})). 
The enhancement of the shock pressure due to the 
increased fluid mass by the factor of $\gh$ is compensated by the 
reduction of pressure due to the fluid pre-acceleration to $\gamma^\prime$. 

We conclude that the blast-wave dynamics should not be strongly changed
by the flash impact on the external medium. More detailed calculations will, 
however, be needed at smaller radii where the effect of the flash radiation on 
the external medium interferes with that of the prompt radiation, increasing 
the pre-acceleration Lorentz factor from $\gamma\gg 1$ to a new 
$\gamma^\prime$.

\subsection{Effect on radiative efficiency}

The deposited heat implies a huge energy per electron ahead of the shock, 
$\gth m_ec^2$. In the region of interest, where $G\gg 1$ and 
$\mu_e  m_p / m_e \gg Z_\pm \gg 1$, one finds
\beq
\gth \approx \gh \frac{\mu_e  m_p}{Z_\pm m_e}\gg 1.
\eeq
When the hot fluid passes through the shock, the thermal Lorentz factor of 
particles increases to $\ginj$ given by \Eq~(\ref{eq:ginj}). Using \Eq~(\ref{eq:rel}),
one obtains
\beq
   \ginj\approx \Gbw\,\frac{\mu_e m_p}{Z_\pm m_e}.
\eeq
This relation shows that all the energy available for dissipation in the 
blast wave ($Z_\pm \ginj m_e c^2 / \mu_e m_p \approx \Gbw c^2$ per unit 
external mass) has been converted into the heat of pairs behind the shock. 
It implies the effective $\epse=1$,
regardless of the efficiency of energy transfer from the ions to 
pairs at the shock front. The high-energy particles behind the 
shock radiate most of their energy and produce radiation beamed within 
angle $\theta\sim\Gbw^{-1}$. Our transfer simulations in \Sect~5 
and analysis in \Sect~6 show that a large fraction of this radiation 
avoids $\gamma$-$\gamma$ absorption and escapes, leading to a high 
radiative efficiency of the blast wave.

%############################################################

\section{Discussion}

\subsection{Mechanism of the GeV flash}

The external shock of the GRB explosion in a dense progenitor wind generates 
a bright GeV flash due to inverse Compton (IC) cooling of the shock-heated 
plasma. We showed that scattering of the prompt MeV radiation streaming 
through the external blast wave is the key mechanism during the main phase 
of the flash, shaping its peak and early decay.

Most MeV photons stream without any interaction, however a small fraction
get scattered, and many of the scattered photons (in particular those
scattered in the external medium {\it ahead} of the forward shock) collide
with other MeV photons and convert to $e^\pm$ pairs.
This leads to a dramatic enhancement of electron
density in the blast wave, by a factor of $Z_\pm\sim 10^4$ at radii 
$R\sim 10^{16}$~cm, and hence a dramatic increase in the 
number of prompt photons scattered in the blast wave.
In addition, the GRB radiation pressure significantly pre-accelerates the 
external medium ahead of the forward shock. 
This effect reduces the strength of the shock and regulates the spectrum 
of its inverse-Compton radiation.

We have examined the inverse-Compton pair-dominated flash 
using a direct radiative transfer simulation. 
As an example, we calculated the flash expected from GRB~080916C,
one of the few brightest GRBs well observed by LAT.
When the reverse shock is relativistic, the dynamics and emission of the
forward shock is indifferent to the precise Lorentz factor of the ejecta $\Gej$; 
only the ejecta power $\Lej$ is important. $\Lej$ can be estimated from the 
observed GRB luminosity assuming a plausible radiative efficiency of the 
prompt emission $\eff<1$. The main remaining parameter  
of the blast wave is the density of the external medium
which depends on the progenitor mass-loss rate $\dot{M}$.
We find that $\dot{M}\approx 10^{-5}M_\odot$~yr$^{-1}$, which is typical 
for Wolf-Rayet stars, gives a GeV flash in striking agreement with observations 
(Figure~\ref{fig_lc}). 
Our results explain the previously puzzling
features of the GeV light curve including the early peak and the long decay. 
The light curve is shaped by the pre-acceleration and 
pair-loading effects; the peak is reached where $\gamma\sim 10$ and 
$Z_\pm\sim 10^4$, when most of the shock energy is emitted in IC photons 
of energy $\EIC\sim(\Gbw/\gpre)^2$~MeV, in the GeV band.

The predicted spectrum in the GeV band has the photon index $\sim -2$
(\Fig~\ref{fig_sp}), which is consistent with observations \citep{lat_2013}.
At the high-energy end, $E\gg 10$~GeV, the spectrum is affected by 
$\gamma$-$\gamma$ absorption.
However, absorption does not strongly suppress the emission even at 
very high energies $E>100$~GeV.
Our analysis in \Sect~6 shows that the main source of  $\gamma$-$\gamma$ 
opacity seen by the GeV photons is the unscattered prompt radiation;
the corresponding optical depth $\tau_{\gamma\gamma}$ is given by 
\Eq~(\ref{eq:pprod:tau3}), 
which is self-regulated to a moderate value comparable to unity.
As a result, we predict escaping gamma-rays at energies 
$E\gg 10$~GeV, up to the TeV range, where the flash can be 
detected by the atmospheric Cherenkov telescopes.
  
When comparing the model with the LAT data we assumed that 
all observed GeV emission comes from the blast wave. In fact, at early 
times, the high-energy tail of the prompt emission may contribute to the 
observed GeV light curve near the peak the flash. Variability detected
at early times provides evidence for such a contribution.
After subtraction of the prompt emission, the true light curve of the GeV 
flash may have a somewhat lower peak, perhaps by a factor $\sim 2$.
Then our best-fit model will need to be revised, resulting in moderate 
changes in $A$, $R_p$, and $\Gamma$.
  
Given the similar light curves of the GeV flashes in many GRBs,
it appears likely that all of them are produced by the same mechanism.
This includes GRB~090510 that was attributed to  the short GRB class,
which is usually associated with a different type of progenitors.
It could be that GRB~090510 is an ``impostor'' and its
progenitor had a significant wind before the explosion. A  wind medium was also suggested 
by \citet{panaitescu_2011} based on the afterglow properties of GRB~090510.
Our preliminary analysis of the GeV flash in GRB~090510 confirms the 
requirement of a high external density at $R\sim 10^{16}$~cm, suggesting
a wind medium.
However, the formal constraints on the density profile in this case are not tight
and will be investigated in a future work. In contrast, the IC flash in GRB~080916C  
requires the density profile to be close to $R^{-2}$; a uniform medium would 
give a GeV light curve much flatter than observed.

\subsection{Approximations used and possible extensions}

From a technical point of view, this paper examined  the coupled problem
of radiative transfer and blast-wave dynamics in a wind medium. The problem 
can be solved exactly from first principles, although in this paper we used 
some approximations. Below we summarize our approximations, discuss the 
accuracy of our results, and outline directions for future work.

(1) We conservatively assumed that the postshock plasma is dominated by 
the {\it thermal} $e^\pm$ population.
This assumption is broadly consistent with observations of collisionless
shocks in the solar system and supernovae, as well as numerical simulations
of relativistic shocks (e.g. \citealt{sironi_2009}).
Our calculations made no additional assumptions concerning
particle acceleration in the shock wave.
The likely presence of a small number of nonthermal particles would 
weakly change the predicted light curve shown in \Fig~\ref{fig_lc}
(as discussed in \Sect~5) except possibly at the earliest stages, before
the peak of the flash. We used the simplest possible approximation where 
the shocked particles acquire the mono-energetic distribution $\delta(\ge-\ginj)$
with $\ginj$ given by \Eq~(\ref{eq:ginj}).  
Detailed future models can use a more realistic distribution, e.g. 
Maxwellian, and include nonthermal particles.

(2) Our calculations had to invoke one phenomenological parameter 
$\varepsilon_e$. The shock wave heats ions and electrons/positrons,
and $\varepsilon_e$ is the fraction of the ion energy that is immediately
(due to collective plasma effects) passed to $e^\pm$.
This parameter is not relevant at the peak of the flash,
however its value can affect the decay after the peak (see \Fig~\ref{fig_lc}).
Future particle-in-cell simulations of pair-loaded shocks may provide 
an estimate for $\epse$. In \Sect~8, we showed that 
the blast wave after the peak of the GeV flash enters a peculiar radiative 
regime which can be described as emission with effective $\epse=1$.
For comparison, \Fig~\ref{fig_lc} also presents the GeV flashes obtained 
with $\epse=0$ and $0.1$; it shows that variations in $\epse$ would have a 
modest effect on the light curve. Comparison with the LAT data in \Fig~\ref{fig_lc}
gives no preference to any $\epse$ at times $\tobs<40$~s. At later times, 
the data favors $\epse>0.1$.
The value of $\epse$ makes a significant difference for the 
flash spectrum at high energies $E\gg 1$~GeV (see \Fig~\ref{fig_sp2}).

(3) The numerical models presented in this paper focused on the main phase of 
the GeV flash and did not include possible IC emission at radii $R>R_1$, where 
$R_1$ is
given by \Eq~(\ref{eq:R1}). In reality, some target photons are available for the 
blast wave even at $R>R_1$ (they are provided by a weaker/softer tail of the 
prompt radiation and by the synchrotron 
emission from the blast wave). The high-energy emission will continue as long 
as the target radiation field is able to drain an interesting fraction of the shock
energy via Compton cooling. Thus, the observed light curve of the GeV flash
can extend to much longer observational times than shown in \Fig~\ref{fig_lc}.
As the radiation density decreases behind the prompt radiation front,
the transition from fast to slow cooling regime will affect the GeV light curve.

(4) We used a simplified ``mechanical'' model for the blast-wave dynamics, 
which treats the shocked gas as one hot body. It is equivalent to assuming 
a flat profile of the fluid Lorentz factor behind the forward shock. 
Future detailed models of GeV flashes will be based on full hydrodynamical
simulations. We found that the light curve of the GeV flash near its maximum
is quite sensitive to small refinements in $\Gamma(R)$, even when these 
refinements are at $\sim 10$\% level. Thus, careful hydrodynamical simulations 
will help improve the accuracy of the explosion reconstruction from the observed 
GeV emission.

(5) We calculated in detail how the scattering of GRB radiation and pair creation 
in the external  medium impacts the forward shock. However, we did not study 
the dynamical effect of pair creation {\it behind} the shock. Many of the photons 
scattered in the external medium propagate into the blast wave and the 
unshocked ejecta, and create pairs there with a rate similar to that ahead of 
the blast wave.  As these pairs are picked up by the relativistic flow, they exert 
a significant drag and heat it. Our preliminary estimates suggest that this effect 
is important for the blast-wave dynamics at early times, and will 
reduce the Lorentz factor $\Gbw$ at small radii $R=10^{15}-10^{16}$~cm.
It can strongly affect the rise of the GeV light curve. 
We defer the full calculation to a future work; it will also include the ``rocket effect'' 
due to anisotropy of IC emission, which will give a push to the blast wave. 
All these effects will likely change the rise to the peak and possibly the peak itself. 
Therefore, we only trust our best-fit value of the wind density parameter 
$A$ within a factor of $\sim 2$.

(6) The full non-linear calculation of radiative transfer is challenging and 
was not completely done in this paper. In particular,  we saw in our simulations
that some rare IC photons (with highest energies and smallest angles) convert 
to $e^\pm$ ahead of the blast wave and 
deposit huge energy and momentum. Thus, the full non-linear 
problem must include the impact of the GeV flash on the external medium,
not only the impact of the prompt radiation. Our analysis of this effect in \Sect~8 
suggests that it does not significantly change the ram pressure in the forward 
shock. However, it has another important implication: it leads to the effective $\epse=1$ 
and enforces the high radiative efficiency of the blast wave. 
Detailed nonlinear simulations of this effect are deferred to a future work.
  
Such simulations will also allow one to explore the following possibility.
The high-energy pairs created in the external medium by the IC flash photons
may not be completely cooled before the shock reaches them and boosts their 
energy even more, producing extremely energetic particles. These particles 
in turn produce more energetic photons, some of which can again convert 
ahead of the shock, injecting new very-high-energy pairs. 
Thus, the following cycle is possible for a small number of particles/photons: 
shock-heating $\rightarrow$ emission of high-energy photons $\rightarrow$ 
photon conversion to $e^\pm$ ahead of the shock $\rightarrow$ shock heating. 
As a result, ultra-high-energy particles could be generated. This bootstrap 
mechanism is similar to ``photon breeding'' proposed by \citet{stern_2006}.

\subsection{Future observational tests}

The predicted peak time of the GeV flash, $\Tp$, depends on the density 
parameter $A$ (\Sect~5.4). Although many bursts detected by LAT have 
$\Tp\ll\Tb$, some may have $\Tp\sim\Tb$. It will be useful to study such bursts 
for the following reason. Our calculations predict that the flash peaks in the GeV 
band, and its emission below 100~MeV is weak and has a hard spectral slope
(\Fig~\ref{fig_sp}). This weak emission can only be seen when the bright prompt 
emission turns off. A flash with $\Tp\sim\Tb$ would still be near its peak at 
$\tobs>\Tb$, and the measurement of its spectrum could be extended below 
100~MeV to test our prediction in this energy band.

Future analysis of the entire sample of LAT bursts will allow one to 
estimate the wind density, the radius and Lorentz factor of the blast-wave, 
and the efficiency of the prompt emission for a number of GRBs.
Our preliminary analysis of the published LAT catalogue of 35 bursts 
\citep{lat_2013} suggests that the density parameter 
$A\sim 10^{11}-10^{12}$~g~cm$^{-1}$
is typical for GRBs with detected GeV flashes. 

The total energy of the GeV flash is roughly  proportional to the 
product of its peak luminosity $L_p$ and its peak time $\Tp$, which 
scales with $A$. We conclude that the flash is likely to be detected in GRBs 
that are bright and exploding in dense stellar winds. This may explain why only  
$\sim 10$\% of GRBs are found to produce strong emission in the GeV band. 
Note also that a relatively low wind density 
is suggested by the analysis of optical afterglows in a sample
of bursts, none of which was detected by LAT \citep{hascoet_2013}.

Observations of the GeV flash determine not only $A$ but 
also $R_p$ and the blast-wave Lorentz factor at $R_p$ (\Sect~5.4). 
In particular, for GRB~080916C we found $R_p\approx 10^{16}$~cm and 
$\Gbw(R_p)\approx  500$.\footnote{This value is in conflict with 
      \citet{abdo_2009}
      who concluded that the GeV source moves with $\Gamma> 890$.
     The discrepancy is explained by the overly pessimistic assumptions of
     \citet{abdo_2009} concerning the angular distribution of the target photons
     (see also \citealt{hascoet_2012}).}
This completely defines the blast wave, and one can extrapolate its dynamics
at later times when the optical and X-ray afterglow emission is observed.
This opens new prospects for understanding afterglow emission of GRBs.
   
The prediction of bright emission above 100~GeV (\Fig~\ref{fig_sp}) can be 
tested with ground-based telescopes. In particular, the High Altitude Water 
Cherenkov telescope \citep{taboada_2013} and the Cherenkov Telescope Array
\citep{inoue_2013} should be able to observe this emission.
We expect that the intrinsic cutoff of the high-energy spectrum at 
$\tobs\simgt 1$~min is above 1~TeV. Then the observed cutoff will be 
shaped by absorption of the flash by the extragalactic background light.

\subsection{Optical flash}

We argued in \Sect~7.2 that the magnetic field in the blast wave may be 
measured through observations of the low-energy (synchrotron) counterpart 
of the GeV flash, in particular in the optical band. A small magnetization 
parameter $\epsB$ would not affect the GeV flash and still 
give bright optical emission which scales as $\epsB^{1/2}$.

For instance $\epsB\sim 10^{-6}$ gives an optical counterpart that 
reaches the peak luminosity comparable to $10^{50}$~erg~s$^{-1}$ in 
$\sim 10(1+z)$~s, followed by a steep decay phase, roughly as $\tobs^{-2}$.
This fast decay is mainly controlled by the quickly decreasing pair-loading 
of the external medium as the blast wave expands past $\sim 10^{16}$~cm.
Most of the shock energy is lost to the fast Compton cooling, and only a 
small fraction is given to the optical synchrotron emission.

The expected optical flash is very similar to the flash observed in 
GRB~990123 \citep{akerlof_1999}.
Note that it reached its peak well before the end of the prompt emission, which 
is consistent with efficient Compton cooling of the flash-producing electrons
\citep{beloborodov_2005}. Unfortunately, GRB~990123 could not be observed 
at high energies (it was too far off axis for EGRET, the only available 
GeV telescope at the time). If our interpretation of the optical flash in GRB~990123
is correct, it should have been accompanied by a bright GeV flash. 

Such double (optical+GeV) flashes may be detected by future simultaneous
observations by {\it Fermi} and optical robotic telescopes at times 
$\tobs\sim(10-100)(1+z)$~s after the burst trigger.
Our calculations predict that the peak of the optical flash 
is slightly delayed compared with the GeV peak and decays faster.
\medskip

When this work was completed, the first detection of a double optical+GeV
flash was reported in GRB~130427A \citep{vestrand_2013}. It confirms the 
predictions of our model. A detailed study of the flash in GRB~130427A
and its implications will be published elsewhere (Vurm et al., in preparation).

\vspace{0.2in}
We are grateful to Nicola Omodei and Sylvain Guiriec for providing 
LAT and GBM data for GRB~080916C.
This work was supported by NSF grant AST-1008334 
and NASA Fermi Cycle 6 grant NNX 13AP246.

%###########################################################

\bibliographystyle{apj}
\bibliography{biblio}

\end{document}